\documentclass[12pt]{article}
\usepackage[polish,english]{babel}
\usepackage[latin1]{inputenc}
\usepackage{graphicx}
\usepackage{times}
\usepackage[T1]{fontenc}

\def\br{\begin{eqnarray}}
\def\er{\end{eqnarray}}
\def\be{\begin{equation}}
\def\ee{\end{equation}}
\def\({\left(}
\def\){\right)}

\def\rlx{\relax\leavevmode}
\def\IR{\rlx\hbox{\rm I\kern-.18em R}}

\def\u2{\mid u\mid^2}

\begin{document}
\begin{titlepage}

\begin{center}
{\bf Properties of some (3+1) dimensional vortex solutions of the $CP^N$ model}
\end{center}

\vspace{.5cm}

\begin{center}
{L. A. Ferreira~$^{\star}$, P. Klimas~$^{\star}$ and W. J. Zakrzewski~$^{\dagger}$}

\vspace{.5 in}
\small

\par \vskip .2in \noindent
$^{(\star)}$Instituto de F\'\i sica de S\~ao Carlos; IFSC/USP;\\
Universidade de S\~ao Paulo  \\ 
Caixa Postal 369, CEP 13560-970, S\~ao Carlos-SP, Brazil\\

\par \vskip .2in \noindent
$^{(\dagger)}$~Department of Mathematical Sciences,\\
 University of Durham, Durham DH1 3LE, U.K.

\end{center}

\begin{abstract}

We construct new classes of vortex-like solutions of the $CP^N$
 model in (3+1) dimensions and discuss some of their properties.
 These  solutions are obtained by generalizing to (3+1) dimensions 
the techniques well established for the two dimensional  $CP^N$ models.
 We show that as the total energy of these solutions is infinite, they   describe evolving vortices and anti-vortices with the
 energy density of some configurations   varying in time.
We also make some further observations about the dynamics of these vortices.

\end{abstract}


\end{titlepage}

\section{Introduction}

In this paper we present new classes of vortex-like solutions of the $CP^N$ model
 \cite{dadda,wojtekbook} in (3+1) dimensions. Our results generalize
 those obtained in our previous  paper \cite{one} where we presented a quite 
large class of exact solutions of $CP^N$ models in ($3+1$) dimensions. 
 These solutions were described by arbitrary functions
of two variables, namely of the combinations $x^1+i\, x^2$ and $x^3+x^0$,
where $x^{\mu}$, $\mu=0,1,2,3$ are the Cartesian coordinates of four
dimensional Minkowski space-time. Then we considered field 
configurations, which for fixed values of
$x^3+x^0$ were holomorphic solutions of the $CP^N$ model
in (2+0) dimensions. The dependence on $x^3+x^0$ was assumed to be in terms
of phase factors ($e^{ik(x^3+x^0)}$). These solutions then described 
straight vortices with waves traveling along them with the speed of light. Solutions of that type were also constructed for an extended version of the Skyrme-Faddeev model \cite{CP1vortex,CPNvortex}. 
 Our previous paper \cite{one}  contained other solutions for which 
the vortices and the waves were in more complicated
interactions with each other.

In this paper we generalize the procedure of \cite{one} 
and generate many more vortex-like solutions and also discuss
solutions which correspond to configurations of parallel vortices
and anti-vortices. Such structures interact with each other and 
our solutions describe this interaction and the resultant dynamics.
 A novelty of the paper is that we generalize to $(3+1)$ dimensions a
 method for constructing solutions which was originally proposed \cite{wojtekbook} 
in the context of the two dimensional $CP^N$ model. Given an holomorphic solution,
 {\it i.e.} a configuration depending only on $x^1+i\, x^2$ and $x^3+x^0$,
 we are able to generate, using a projection operator, solutions
 depending on $x^1+i\, x^2$, $x^3+x^0$ and also $x^1-i\, x^2$.

As our solutions describe vortices their total energy is infinite so to 
compare various configurations of vortices it is convenient to talk
of energy density or energy per unit length. Then, as we discuss in this paper
interesting phenomena that can take place - the energy per unit length
can stay constant, be periodic in time or even grow with time.
At first sight this may seem surprising but, in fact, this is not
in contradiction of any principles, as the total energy remains infinite
and so is ``constant'' ({\it i.e.} does not change). This observation complements
 the observation of our previous paper \cite{one} in which we pointed out that
 although the energy per unit length of various parallel vortex configurations can 
depend on the distance between them the vortices would still remain at rest.

The paper is organized as follows. In the next section, for completeness,
we introduce our notation and recall some basic properties of the $CP^{N}$ models
and of their classical solutions in (2+0) dimensions.

The next section presents our solutions and the following one discusses 
some properties of these solutions. We finish the paper with a short section
presenting our conclusions and further remarks.

\section{General remarks about the $CP^{N}$ model}

The $CP^{N}$ model in ($3+1$) dimensional Minkowski
space-time is defined in terms of its Lagrangian density 
\begin{eqnarray}\label{lagr1}
\mathcal{L}=M^2(D_{\mu}{\cal Z})^{\dagger}D^{\mu}{\cal Z}, \qquad
{\cal Z}^{\dagger}\cdot {\cal Z}=1,
\label{cpnlagrangian}
\end{eqnarray}
where $M^2$ is a constant with the dimension of mass,
${\cal Z}=({\cal Z}_1,\ldots,{\cal Z}_{N+1})\in \mathcal{C}^{N+1}$ and
it satisfies the 
constraint ${\cal Z}^{\dagger}\cdot {\cal Z}=1.$. The covariant
derivative $D_{\mu}$ 
acts on any $N$ component vector $\Psi$ and so also on ${\cal Z}$, according to 
$$D_{\mu}{\Psi}=\partial_{\mu}{\Psi}-({\cal Z}^{\dagger}\cdot
\partial_{\mu}{\cal Z}){\Psi }. 
$$
 The index $\mu$ runs here
 over the set $\mu=\{0,1,2,3\}$ and the Minkowski metric
is 
(+,-,-,-). The Lagrangian
 (\ref{cpnlagrangian}) is invariant under the global transformation \\
 ${\cal Z}\rightarrow U\,{\cal Z}$, with $U$ being a $(N+1)\times
 (N+1)$ unitary  matrix.  One of the advantages of the
 ${\cal Z}$ parametrization is that it makes this $U(N+1)$ symmetry
 explicit \cite{dadda,wojtekbook}. It is also convenient to use the 
`un-normalized' vectors $\hat Z$ with components $\hat Z_i$. Then
\begin{equation}
\label{unnormalised}
{\cal Z}\,=\,\frac{\hat Z}{\sqrt{\hat Z^{\dagger}\cdot \hat Z}},
\end{equation}
where the dot product involves the summation over all ($N+1$) components of $\hat Z$.

Sometimes, exploiting the full projective space symmetry of the model, we set $u=\frac{\hat Z}{\hat Z_{N+1}}$
 and so use the
 parametrization  
\begin{eqnarray}
{\cal Z}=\frac{(1,u_1,\ldots,u_N)}{\sqrt{1+|u_1|^2+\ldots+|u_N|^2}}.
\label{udef}
\end{eqnarray}
The
$u$-field parametrization does not make the $U(N+1)$ symmetry
explicit but it has the advantage that it brings out the real degrees
of the freedom of the model.
In terms of $u_i$'s the Lagrangian density (\ref{lagr1}) takes the form
\begin{eqnarray}\label{lagr2}
\mathcal{L}=\frac{4M^2}{(1+u^{\dagger}\cdot
  u)^2}\left[(1+u^{\dagger}\cdot
  u)\partial^{\mu}u^{\dagger}\cdot\partial_{\mu}u-(\partial^{\mu}u^{\dagger}\cdot
  u)(u^{\dagger}\cdot \partial_{\mu}u)\right]. 
\end{eqnarray}
The classical solutions of the model are given by the $N$ Euler-Lagrange equations which take 
the form: 
\begin{eqnarray}\label{cpn_equation}
(1+u^{\dagger}\cdot
  u)\,\partial^{\mu}\partial_{\mu}u_k-2(u^{\dagger}\cdot
  \partial^{\mu}u)\,\partial_{\mu}u_k=0. 
\end{eqnarray}
  The simplest $CP^1$ case is given  by
one function $u$: ${\cal Z}=\frac{(1,\,u)}{\sqrt{1+|u|^2}}$.

\section{Some  solutions}

In this paper we shall use the notation of \cite{one} {\it i.e.} we define
\begin{equation}
z\equiv x^1+i\,\varepsilon_1 x^2,\qquad {\bar z}\equiv x^1-i\,\varepsilon_1 x^2,
 \qquad y_{\pm}\equiv x^3\pm\epsilon_2\,x^0
\label{zydef}
\end{equation}
with $\varepsilon_a=\pm 1$, $a=1,2$.

It is easy to check that any set of functions $\hat Z_k$ and so $u_k$ that depend on
coordinates $x^{\mu}$ in a special way, namely  
\begin{equation}u_k=u_k(z,y_{+})
\label{gensolution}
\end{equation}
is a solution of the system of equations (\ref{cpn_equation}). The Minkowski
metric in the coordinates (\ref{zydef}) becomes $ds^2=-dz\,d{\bar
  z}-dy_{+}\,dy_{-}$. It then follows that (\ref{gensolution}) 
 satisfies simultaneously $\partial^{\mu}\partial_{\mu}u_i=0$
 and $\partial^{\mu}u_i\partial_{\mu}u_j=0$ for all $i,\,j = 1,\ldots,
 N$. Hence this class of solutions is quite large. 

However, these are not the only solutions we can construct very easily.
In fact, we can exploit the construction \cite{wojtekbook}  of the solutions of the $CP^{N}$ model in (2+0)
dimensions (for $N>1$) to obtain further solutions. 
To do this we recall the construction in (2+0) dimensions:

First we define a Gramm-Schmidt orthogonalising operator $P_z$ by its action on any vector 
$f\in \mathcal{C}^{N+1}$, namely
\begin{equation}
\label{ortho}
P_zf \,=\, \partial_z f\,-\, f \, \frac{f^{\dagger}\cdot \partial_zf}{\vert f\vert^2}.
\end{equation}

Then, if we take $f=f(z)$ and consider $\hat Z=f(z)$ the corresponding $u$ solves the equations  (\ref{cpn_equation}). Note that as $f(z)$ does not depend on $y_{\pm}$ we have a solution of the $CP^{N}$ 
model in (2+0) and in (3+1) dimensions. However, as is well known, (see {\it e.g.} 
\cite{wojtekbook} and the references
therein) 
\begin{equation}
\label{higher}
\hat Z\,=\,P_z f(z)
\end{equation} 
defines further $u$'s which also solve  (\ref{cpn_equation}) in (2+0) dimensions. But, as the expression
for $u$ does not depend on $y_{\pm}$ these functions also solve the equations (\ref{cpn_equation}) in (3+1) dimensions. This procedure can then be repeated, namely we can take
\begin{equation}
\label{higherstill}
\hat Z\,=\,P_z^{k} f(z),
\end{equation}
where $P_z^{k}f=P_z(P_z^{k-1}f)$.

To have more general solutions we observe that, like in \cite{one}, we can make
the coefficients of $z$ in the original $f(z)$ to be functions of one of $y_{\pm}$,
 say, $y_+$.
As $y_+$ is real the operation of applying $P_z$ operator does not introduce the other $y_{\pm}$,
{\it i.e.} $y_-$, and so the corresponding $\hat Z$ and so $u$ give us further solutions of the 
equations (\ref{cpn_equation}) in (3+1) dimensions. This way for $N>1$ we can have 
holomorphic solutions and also `mixed'
solutions. 

They are given, respectively, by
\begin{equation}
\label{holo}
u_k(z,y_+)=\frac{f_k(z,y_+)}{f_{N+1}(z,y_+)}
\end{equation}
and 
\begin{eqnarray}
\label{mixed}
\label{sol_v}
u_k(z,\bar{z},y_+)\equiv \frac{P_z^lf_k}{P_z f_{N+1}}.
\end{eqnarray}  

Note that like in the (2+0) case the last (as we take larger $l$) nonvanishing solution would be antiholomorphic.
Then the corresponding $u_k$ will be functions of only $\bar{z}$ and $y_+$.

\subsection{Some properties of our solutions}
Let us first discuss briefly some quantities which we will use in the discussion of
various properties 
of our solutions.

\subsubsection{The energy of the solutions}

The Hamiltonian density of the 
$CP^N$ model, when written in coordinates
 ($z$, $\bar{z}$, $y_+$, $y_-$), takes the form
\begin{eqnarray}
\mathcal{H}=\mathcal{H}^{(1)}+\mathcal{H}^{(2)},
\end{eqnarray}
where
\begin{eqnarray}
&&\mathcal{H}^{(1)}=\frac{8M^2}{(1+u^{\dagger} \cdot
  u)^2}\left[\partial_{\bar{z}} u^{\dagger}\cdot \Delta^2
  \cdot\partial_z u+\partial_z u^{\dagger}\cdot \Delta^2
  \cdot\partial_{\bar{z}} u\right]
  \label{hamiltonian1}\\
 && \mathcal{H}^{(2)}=\frac{8M^2}{(1+u^{\dagger} \cdot
  u)^2}\left[\partial_{+} u^{\dagger}\cdot \Delta^2
  \cdot\partial_{+} u+\partial_{-} u^{\dagger}\cdot \Delta^2
  \cdot\partial_{-} u\right]
\label{hamiltonian2}
\end{eqnarray}
and $\Delta^2_{ij}\equiv(1+u^{\dagger}
\cdot u)\delta_{ij}-u_iu_j^*$.

For solutions depending on $y_{+}$ {\it i.e.} described by
 $u_k(z,\bar{z},y_+)$ the part of the Hamiltonian density (\ref{hamiltonian2}) 
that contains $\partial_-$ drops out.
For the holomorphic solutions the second part of (\ref{hamiltonian1}) also drops out. 
 For the `mixed' solutions 
described by  (\ref{mixed}) both parts of (\ref{hamiltonian1}) are nonzero.

Note that as our solutions depend on variables $x^0$ and $x^3$ only through the combination $y_+$ it is useful
to define  the concept of energy per unit length which involves the
integration over  $x^1$ and $x^2$ ({\it i.e.} over
 the plane perpendicular to the $x^3$  axis). This gives us 
$$
\mathcal{E}=\int_{R^2}dx^1\,dx^2\,\mathcal{H}=8\pi
 M^2\left[\mathcal{I}^{(1)}+\mathcal{I}^{(2)}\right],
$$
where 
$$
\mathcal{I}^{(a)}\equiv\frac{1}{8\pi M^2}\int_{R^2}dx^1\,dx^2\,\mathcal{H}^{(a)},\qquad a=1,2.
$$

\subsubsection{The topological charge}
As we are working with vortex configurations it is important to introduce the 
two-dimensional topological charge defined
 by the integral
\begin{eqnarray}
\label{topcharge}
Q_{\rm top}=\int_{R^2}dx^1dx^2\rho_{top}\label{Q}
\end{eqnarray}
whose density is given by
\begin{eqnarray}
\rho_{\rm top}&=&\frac{1}{\pi}\varepsilon_{ij}(D_{i}\mathcal{Z})^{\dagger}\cdot(D_{j}\mathcal{Z})=\frac{1}{\pi}\varepsilon_{ij}\frac{\partial_{i}u^{\dagger}\cdot \Delta^2\cdot\partial_{j}u}{(1+u^{\dagger}\cdot u)^2}\nonumber=\\&=&\frac{1}{\pi}\frac{\partial_{\bar z}u^{\dagger}\cdot \Delta^2\cdot\partial_{z}u-\partial_{z}u^{\dagger}\cdot \Delta^2\cdot\partial_{\bar z}u}{(1+u^{\dagger}\cdot u)^2}.\label{density}
\end{eqnarray}
The indices $i$ and $j$ here only
take two values $\{1,2\}$. It is easy to see that for the holomorphic solution $Q_{\rm top}=\mathcal{I}^{(1)}
$.

\section{Vortex solutions of the $CP^{N}$ model and some of their properties}

In \cite{one} we studied  some general classes of 
 solutions of the $CP^1$ model. Here, first of all,
 we concentrate our attention on two classes 
of holomorphic solutions of the  $CP^1$ model 
and then look in some detail at the $CP^2$ model concentrating our attention this time on
`mixed'
solutions (\ref{mixed}).

\subsection{$CP^1$ solutions}

In the $CP^1$ model we have two functions $f_1$ and $f_2$ and in our discussion we can take their ratio 
$u=\frac{f_1}{f_2}$.

Let us first consider the case when all
the dependence on  $y_+$ is in 
 the form of phase factors
 $e^{ik_i
y_+}$ where $k_i$ are
 constant. Many interesting features are observed for
the configurations given by
\begin{eqnarray}
f_1(z,y_+)=z^2+a_1\,z\,e^{ik_1y_+},\qquad f_2(z,y_+)=a_2\,z+a_3\,e^{ik_2y_+},\label{sys1}
\end{eqnarray}
where we have assumed, for simplicity, that all three parameters $a_1$, $a_2$ and $a_3$ are real. The generalization 
to their complex values does not bring anything new to the problem.
 
The holomorphic solution $u$ is then of the form
\begin{eqnarray}\label{sol}
u(z,y_+)=z\frac{z+a_1\,e^{ik_1y_+}}{a_2\,z+a_3\,e^{ik_2y_+}}.
\end{eqnarray}

The zeros of denominator do not lead to the
singularities in the energy density as both integrals $\mathcal{I}^{(1)}$ and $\mathcal{I}^{(2)}$
are invariant with respect to the inversion $u\rightarrow \frac{1}{u}$. 

Next we look in detail at various special cases of this solution (\ref{sol}).

\subsubsection{The tube solution}
First we consider the case of $a_1=a_2=0$. In this case the field configuration becomes
\be
\label{tubesol}
u=\frac{z^2}{a_3}e^{-ik_2y_+}.
\ee

It is easy to convince oneself that this field configuration describes  
 a vortex with waves traveling along it with the speed of light. The profile of the energy density is independent of  $y_+$. It has a maximum at a ring of radius $r_0$ which satisfies $r_1<r_0<r_2$, where $r_1=\sqrt{\frac{|a_3|}{\sqrt{3}}}$ is the radius of the circle at which the Hamiltonian density $\mathcal{H}^{(1)}$ has a maximum, and $r_2=\sqrt{|a_3|}$ corresponds to the radius of the circle at which $\mathcal{H}^{(2)}$ has a maximum. The radius $r_0$ depends on $a_3$ and $k_2$. For $k_2\rightarrow 0$ it tends to $r_1$ and for $k_2\rightarrow \pm\infty$ it tends to $r_2$. The integral $\mathcal{I}^{(1)}$ describes the topological charge of the vortex which for the solution considered here is 
$$
\mathcal{I}^{(1)}=\frac{1}{\pi}\int_{R^2}dx^1dx^2\frac{4a_3^2|z|^2}{(a_3^2+|z|^4)^2}=2.
$$

The contribution to the energy per unit length that comes 
from the traveling waves can be also calculated explicitly. We find
$$
\mathcal{I}^{(2)}=\frac{1}{\pi}\int_{R^2}dx^1dx^2\,
k_2^2\frac{a_3^2|z|^4}{(a_3^2+|z|^4)^2}=\frac{\pi}{4}k_2^2|a_3|.
$$

\begin{figure}
\begin{center}
\includegraphics[width=0.5\textwidth]{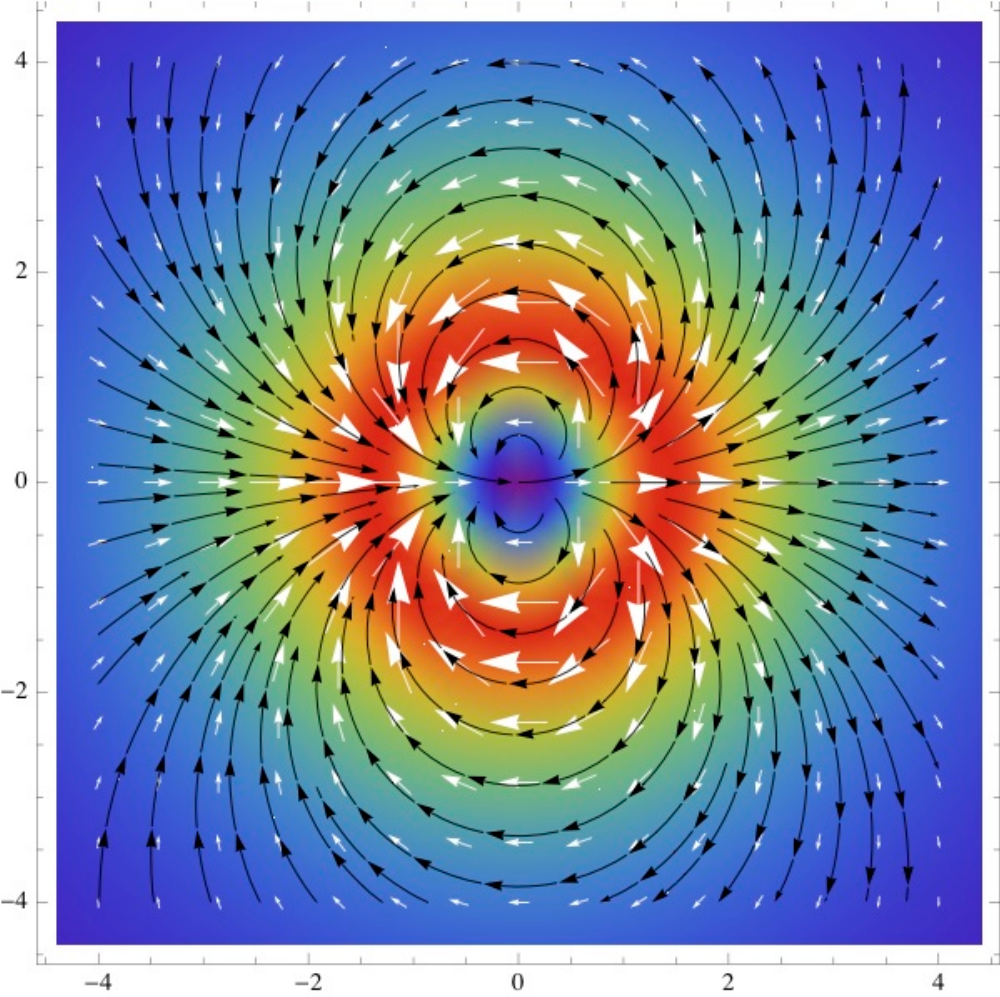}\includegraphics[width=0.5\textwidth]{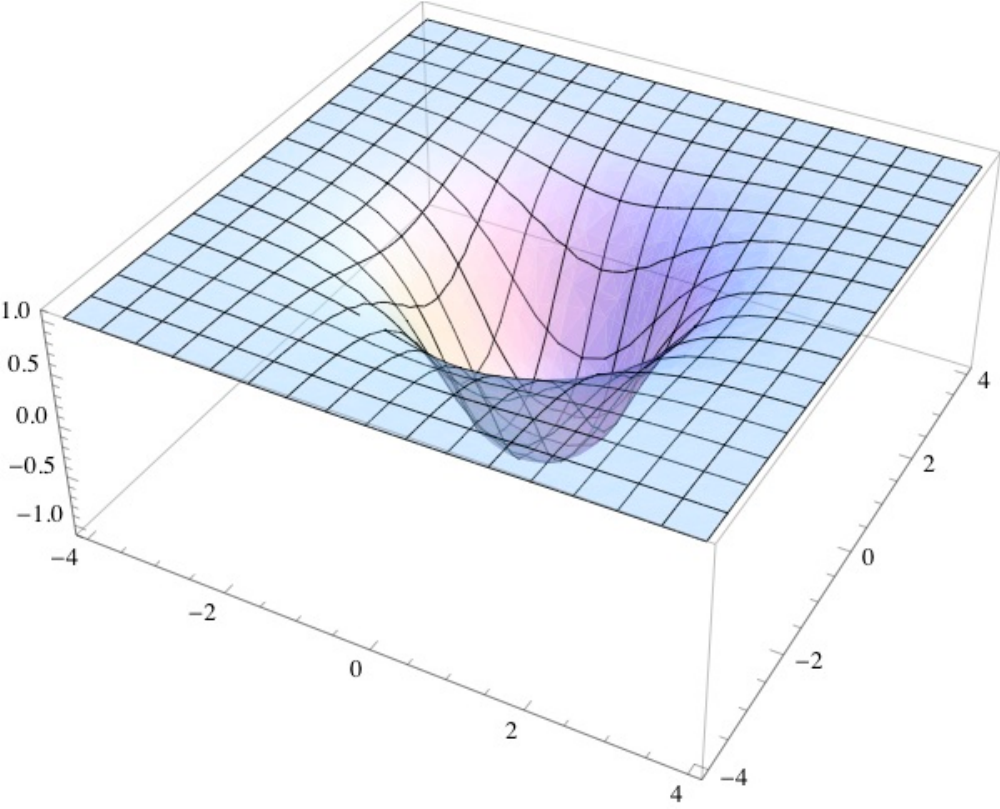}
\caption{The tube solution. The part $(n^1,n^2)$ (left) and the component $n^3$ (right) of the isovector $\vec n$ for $a_1=0$, $a_2=0$, $a_3=2$, $x^0=0$, $x^3=0$ and $k_2=2$. The minimal value $n^3=-1$ occurs at the point $x^1=0$ and $x^2=0$.}
\label{tube_isovector}
\end{center}
\end{figure}

\begin{figure}
\begin{center}
\includegraphics[width=0.5\textwidth]{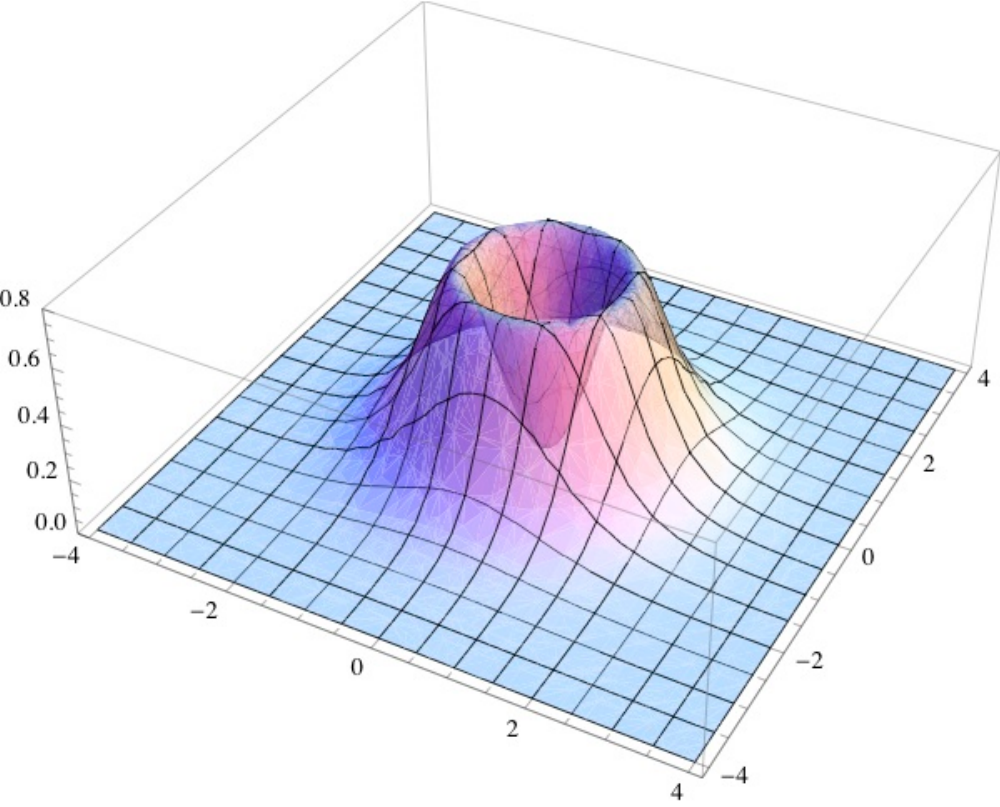}\includegraphics[width=0.5\textwidth]{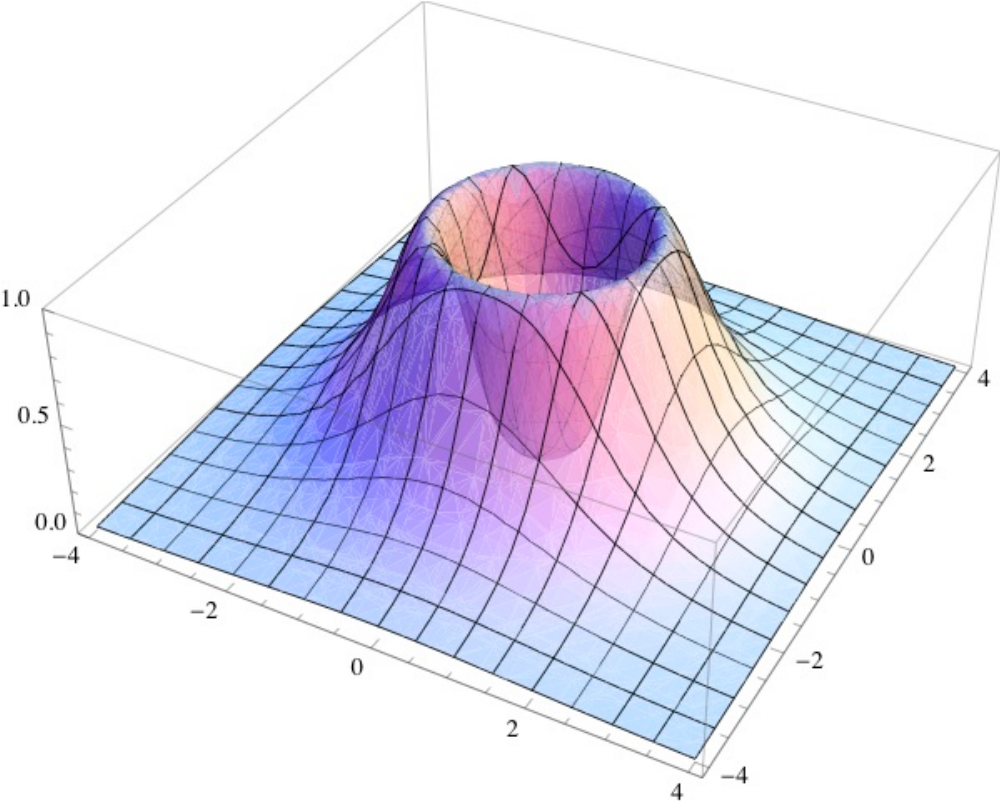}
\caption{The energy density of the tube solution - the topological part (left) and the wave part (right).  Here $a_1=0$, $a_2=0$, $a_3=2$, $x^0=0$, $x^3=0$ and $k_2=2$.}
\label{tube_ener}
\end{center}
\end{figure}

A modification of a solution of this type had been already studied in \cite{one}. An example of such a solution 
is shown in Fig \ref{tube_isovector},  where we plot  the components  of the isovector
\be
\label{isovector}
\vec n=\frac{1}{1+|u|^2}\left(u+u^*,-i(u-u^*),|u|^2-1\right)
\ee
which depend on $y_+$. As $y_+$ changes the images in Fig.\ref{tube_isovector} rotate. 
In Fig. \ref{tube_ener},  we plot the two contributions, topological and wave, of the energy density on the solution (\ref{tubesol}).

\subsubsection{The spiral solution}

A less trivial but still a very simple solution is obtained from (\ref{sol})
 by putting $a_3=0$, and so $u$ is given by 
\be
\label{spiralsol}
u=\frac{1}{a_2}(z+a_1e^{ik_1y_+}).
\ee

In this case the integrals $\mathcal{I}^{(1)}$ and $\mathcal{I}^{(2)}$ can be calculated explicitly. They 
take the values 
\begin{eqnarray}
&&\mathcal{I}^{(1)}=\frac{1}{\pi}\int_{R^2}dx^1dx^2\frac{1}{a_2^2}\frac{1}{(1+|u|^2)^2}=1,\label{int1}
\\
&&\mathcal{I}^{(2)}=\frac{1}{\pi}\int_{R^2}dx^1dx^2\frac{1}{a_2^2}\frac{a_1^2 k_1^2}{(1+|u|^2)^2}=a_1^2 k_1^2.\label{int2}
\end{eqnarray}

\begin{figure}
\begin{center}
\includegraphics[width=0.5\textwidth]{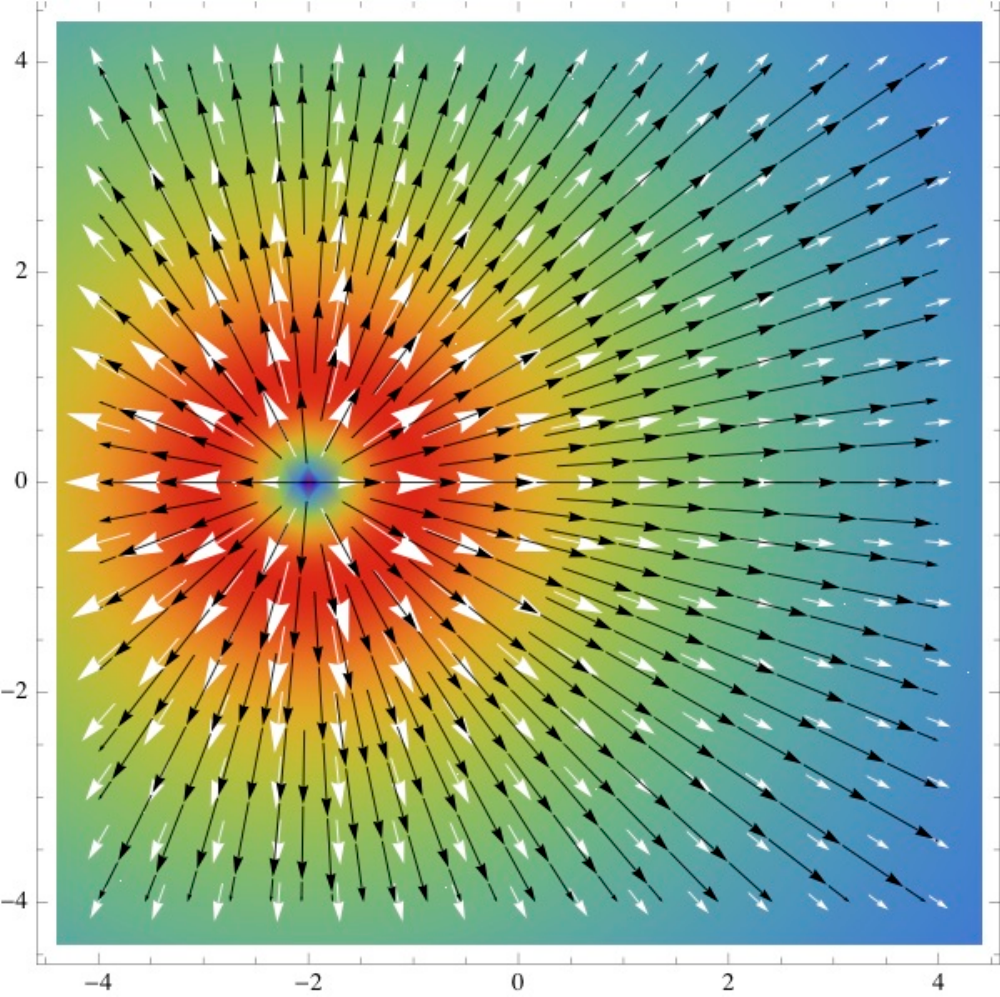}\includegraphics[width=0.5\textwidth]{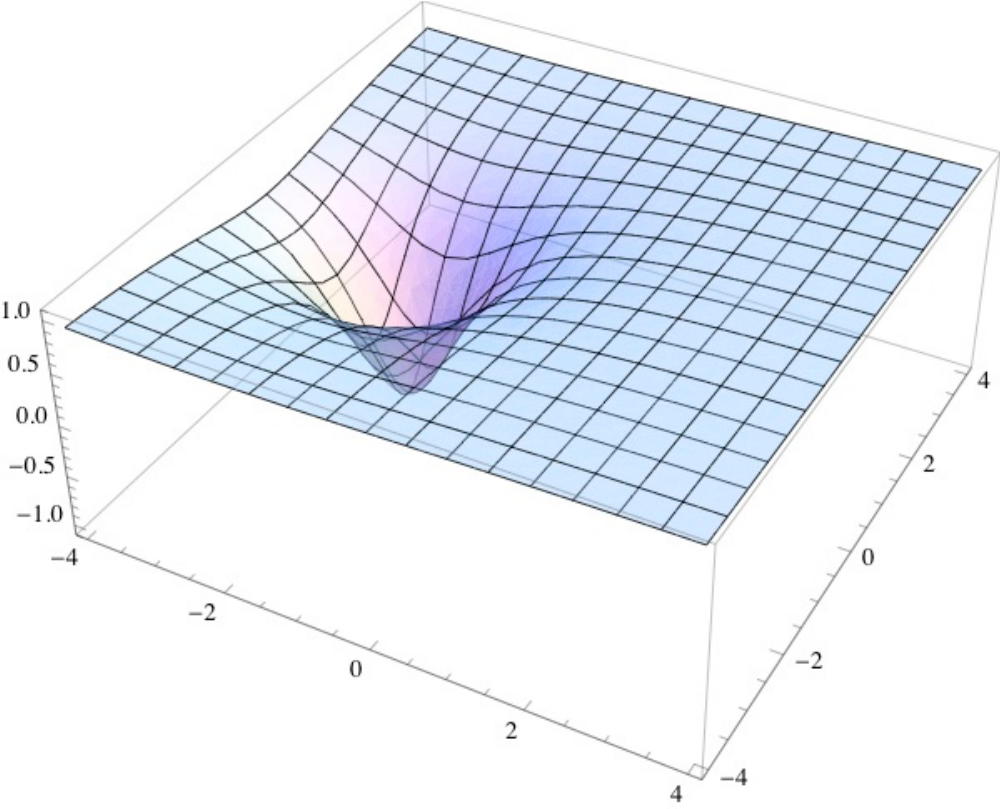}
\caption{The spiral solution. The part $(n^1,n^2)$ (left) and the component $n^3$ (right) of the isovector $\vec n$ for $a_1=2$, $a_2=1$, $a_3=0$, $x^0=0$, $x^3=0$ and $k_1=1$. The minimal value $n^3=-1$ occurs at the point $x^1=-a_1\cos{(k_1 y_+)}$ and $x^2=-a_1\sin{(k_1 y_+)}$; here $x^1=-2$, $x^2=0$.}
\label{spiral_isovector}
\end{center}
\end{figure}

In Fig. \ref{spiral_isovector} we plot the components of the
 isovector (\ref{isovector}) for the solution (\ref{spiralsol}).
In order to analyze the energy density let us introduce  the parameterization
 $z=r e^{i\varphi}$. Then 
$$
|u|^2=\frac{1}{a_2^2}\left[r^2+a_1^2-2a_1 r \cos(\varphi-k_1y_+-\pi)\right]
$$
We note that the energy per unit length (${\cal H}$ integrated over the
 $x^1\, x^2$ plane) does not depend on $a_2$ or $y_+$, whereas the energy
 density ${\cal H}$ does. The maximum of the energy density ($|u|^2=0$) is
 located at $r=|a_1|$ and $\varphi=k_1y_++\pi$.
 The curve $(a_1\cos{(k_1y_++\pi)},a_1\sin{(k_1y_+ +\pi)},y_+)$ that
 joins the points at which the energy density has a local maximum is a spiral.
 On this spiral not only $\mathcal{H}$ has a maximum but so do also both its 
 contributions $\mathcal{H}^{(1)}$ and $\mathcal{H}^{(2)}$. As $y_+=x^3+x^0$, we
 note that the spiral rotates around the $x^3$
axis with the speed of light. The only effect 
of the dependence on $y_+$ is the rotation of the energy density.
 Thus the energy per unit length calculated for {\it e.g.} $y_+=0$ is also
 valid for other values of the variable $y_+$. 

\subsubsection{The general case (\ref{sol})}

For general values of $a_1$, $a_2$ and $a_3$ the expressions for the contributions to the energy become rather
complicated. We can write them as 
\begin{eqnarray}
\mathcal{I}^{(1)}=\frac{1}{\pi}\int_{R^2}dx^1dx^2\frac{A}{C^2},\qquad \mathcal{I}^{(2)}=\frac{1}{\pi}\int_{R^2}dx^1dx^2\frac{B}{C^2}\label{integrals2},
\end{eqnarray}
where the expressions for $A$, $B$ and $C$ take the form (written 
 in cylindrical coordinates $(r,\varphi,y_+)$ with $z=re^{i\varphi}$)
\begin{eqnarray}
A&=&a_1^2a_3^2+4a_3^2r^2+a_2^2r^4+2a_1a_2a_3r^2\cos{[2\varphi-(k_1+k_2)y_+]}\nonumber\\
&+&4a_1a_3^2r \cos{(\varphi-k_1y_+)}+4a_2a_3r^3\cos{(\varphi-k_2y_+)}
\label{A}\\
B&=&r^2a_1^2a_3^2(k_1-k_2)^2+r^4(a_1^2a_2^2k_1^2+a_3^2k_2^2)\nonumber\\
&-&2a_1a_3^2(k_1-k_2)k_2r^3\cos{[\varphi-k_1y_+]}\nonumber\\
&+&2a_1^2a_2a_3(k_1-k_2)k_1r^3\cos{[\varphi-k_2y_+]}\nonumber\\
&-&2a_1a_2a_3k_1k_2r^4\cos{[(k_1-k_2)y_+]}\label{B}
\\
C&=&r^2\left[r^2+2a_1r\cos{[\varphi-k_1y_+]}+a_1^2\right]\nonumber\\&+&a_2^2\left[r^2+2\frac{a_3}{a_2}r\cos{[\varphi-k_2y_+]}+\left(\frac{a_3}{a_2}\right)^2\right]
\label{C}.
\end{eqnarray}

\begin{figure}
\begin{center}
\includegraphics[width=0.5\textwidth]{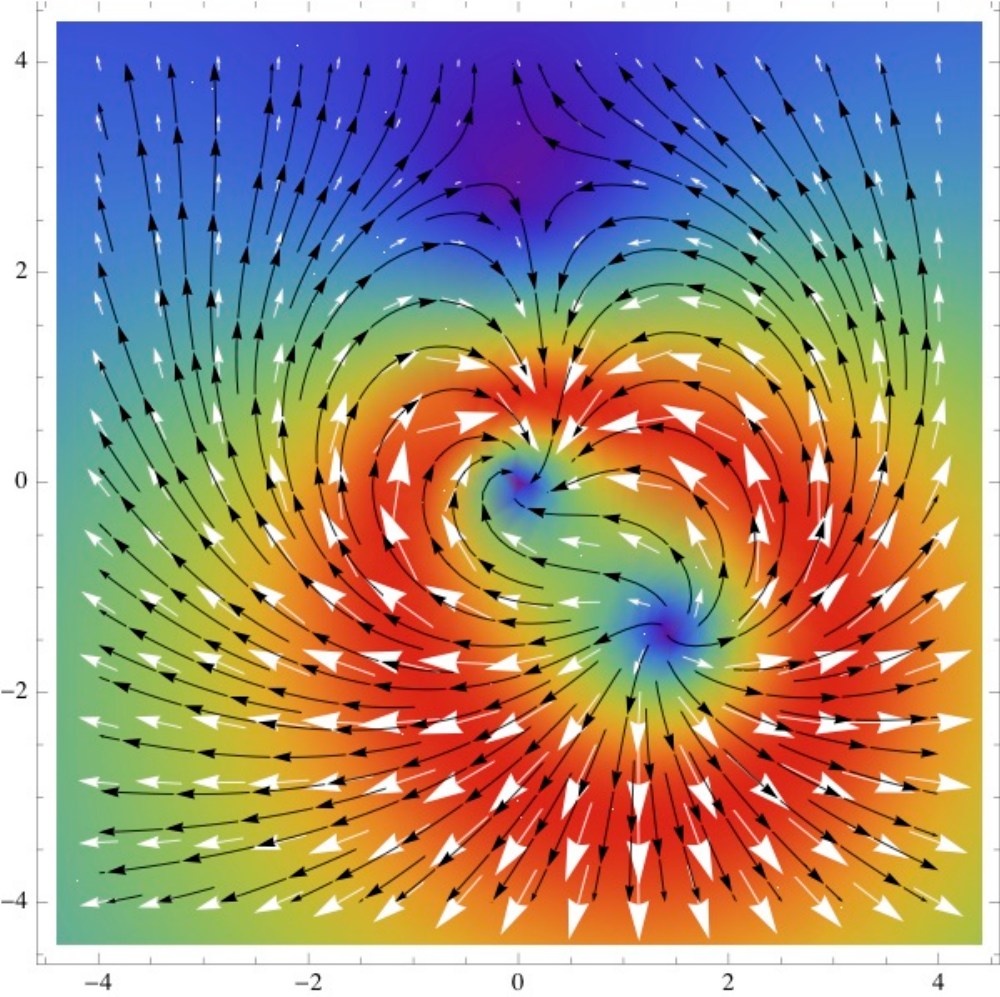}\includegraphics[width=0.5\textwidth]{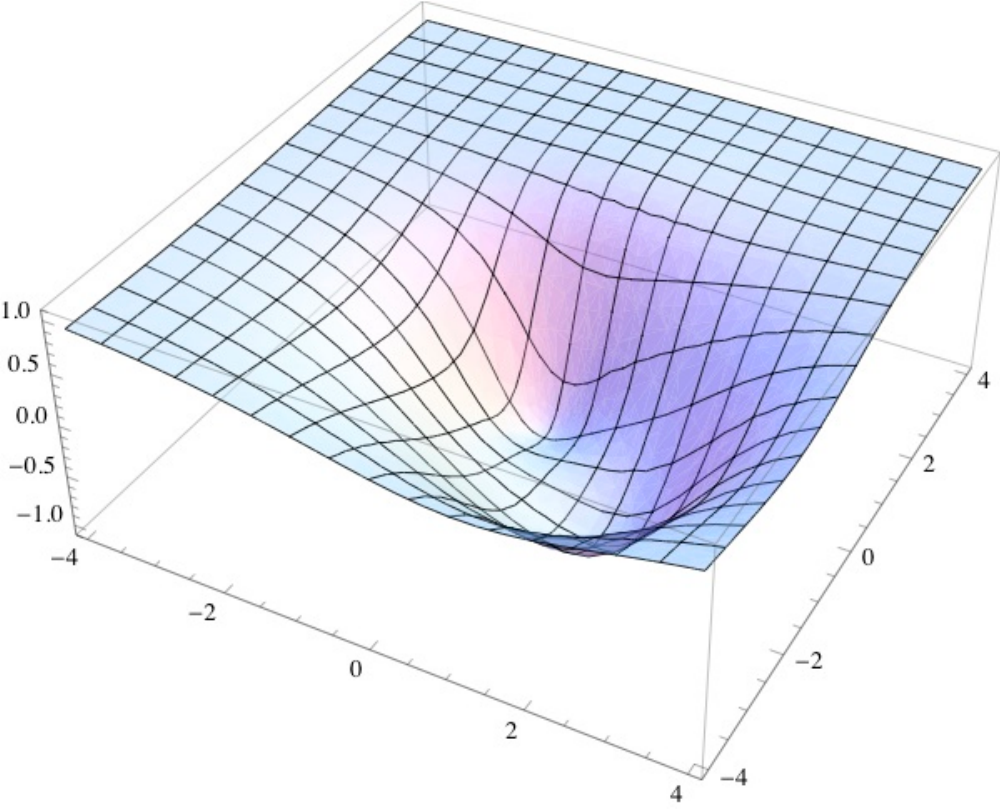}
\caption{The $CP^1$ solution with all $a_k\neq 0$. The part $(n^1,n^2)$ (left) and the component $n^3$ (right) of isovector $\vec n$ for $a_1=2$, $a_2=1$, $a_3=3$, $x^0=3\pi/4$, $x^3=0$, $k_1=1$ and $k_2=2$.}
\label{gen_isovector}
\end{center}
\end{figure}

\begin{figure}
\begin{center}
\includegraphics[width=0.5\textwidth]{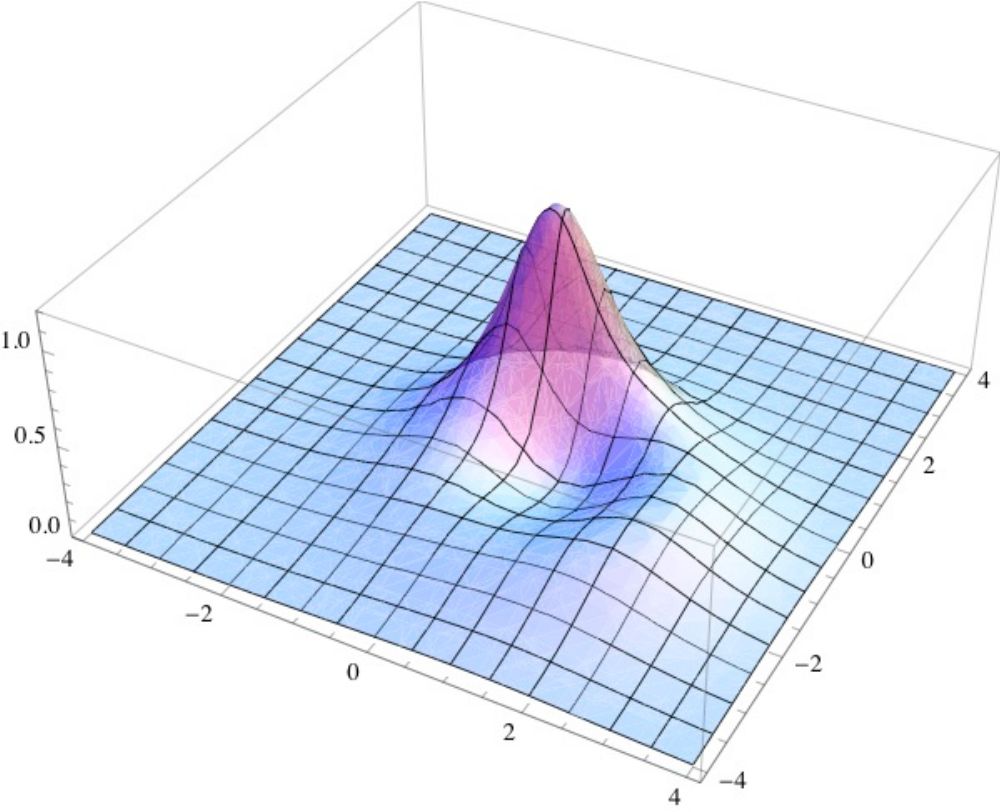}\includegraphics[width=0.5\textwidth]{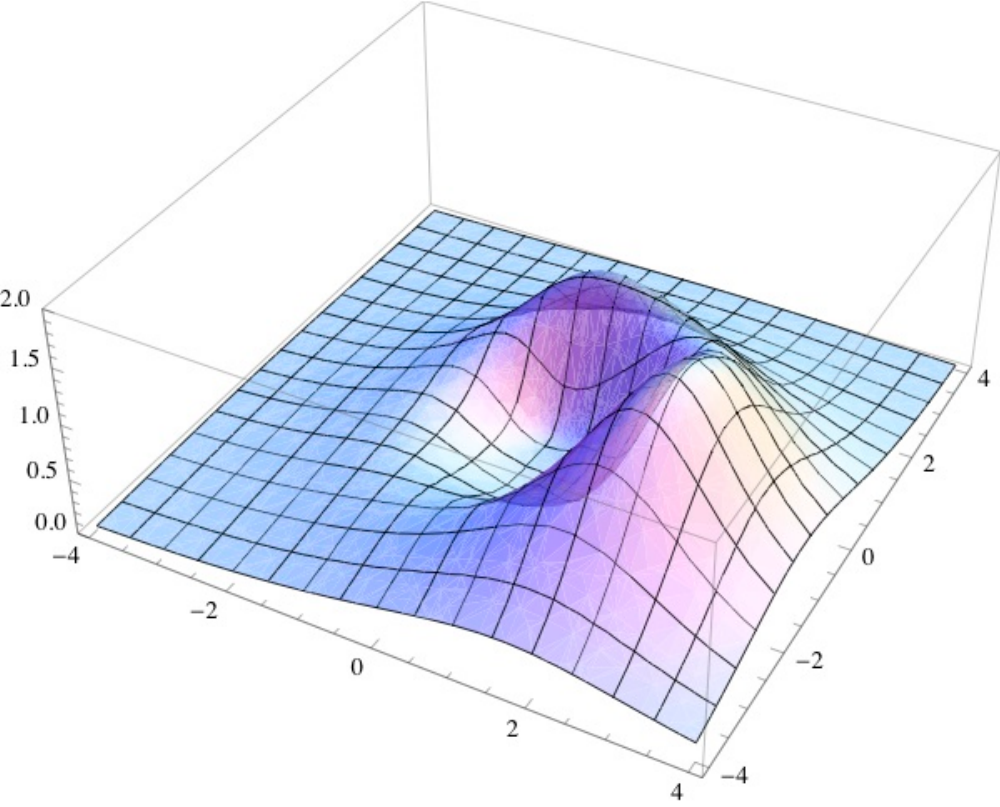}
\caption{The energy density of the $CP^1$ solution with all $a_k\neq 0$ - the topological part (left) and the wave part (right).  Here $a_1=2$, $a_2=1$, $a_3=3$, $x^0=3\pi/4$, $x^3=0$, $k_1=1$ and $k_2=2$.}
\label{gen_ener}
\end{center}
\end{figure}

To fully analyse these expressions requires numerical work.
 In Figs. \ref{gen_isovector} and \ref{gen_ener} we present the plots of 
the isovector (\ref{isovector}) as well as of the energy densities for a
 particular example of the above solution. However, even for a general configuration,
 it is possible to make a few analytical observations:
\begin{itemize}
\item {Rotations:}

Note that the energy per unit length depends on $y_+$ through periodic functions,
 involving four frequencies, namely $k_1$, $k_2$ and $k_1\pm k_2$.
 However, one can isolate  four situations where only one frequency is relevant 
and the time evolution reduces to a rotation around the $x^3$-axis.
 In such cases, $A$, $B$ and $C$ depend on $\varphi$ and $y_{+}$ only
 through the combination $\varphi - \omega\, y_{+}$, and the four possibilities when this happens
 are:  

\begin{enumerate}
\item
$k_1=k_2\equiv k$ and $\omega = k$
\item
$a_1=0$  and $\omega = k_2$
\item
$a_2=0$ and $\omega = k_1$
\item
$a_3=0$ and $\omega = k_1$
\end{enumerate}

Note that the spiral solution (\ref{spiralsol}) belongs to the last case
 and the tube solution (\ref{tubesol}) corresponds to the case when none of the
 frequencies matters. 

\item {Singularity}

The solution (\ref{sol}) exhibits an interesting property when
$a_1=\frac{a_3}{a_2}$.  Indeed, in this case it reduces to $u=z/a_2$ whenever
 $(k_2-k_1)y_{+}=2\pi n$, with $n$ integer. The case $k_1=k_2$ is not interesting 
since it leads to a solution independent of $y_{+}$. However, for $k_1\neq k_2$ the
 solutions change their properties, including the two dimensional topological charge
 (\ref{topcharge}),  whenever $y_+=\xi_n\equiv\frac{2\pi n}{k_2-k_1}$. For those 
special values of $y_{+}$  the quantities (\ref{A})-(\ref{B}) become
$$
A=a_2^2|\vec r-\vec r_n|^4,\qquad B=a_3^2(k_1-k_2)^2r^2|\vec r-\vec r_n|^2,\qquad C=(r^2+a_2^2)|\vec r-\vec r_n|^2
$$
where $\vec r$ and $\vec r_n$ are two-component vectors: $\vec r\rightarrow (x,y)$, and $\vec r_n\rightarrow (x_n,y_n)$, with 
\begin{eqnarray}
x_n=\frac{a_3}{a_2}\cos{(k_1\xi_n+\pi)}, \qquad y_n=\frac{a_3}{a_2}\sin{(k_1\xi_n+\pi)}.
\end{eqnarray}
The expression $|\vec r-\vec r_n|^2$ then becomes
\begin{eqnarray}
|\vec r-\vec r_n|^2&=&(x-x_n)^2+(y-y_n)^2\nonumber\\&=&r^2-2\frac{a_3}{a_2}r\cos{(\varphi -k_1\xi_n-\pi)}+\left(\frac{a_3}{a_2}\right)^2.
\end{eqnarray}

The cancelation changes the degree of polynomials of variable $z$ which causes 
the topological charge to jump from  $Q_{\rm top}=2$ down to $Q_{\rm top}=1$.  The new topological charge is then given by the integral $\mathcal{I}^{(1)}$
\begin{eqnarray}
\label{new}
Q_{\rm{top}}\equiv\mathcal{I}^{(1)}=\frac{1}{\pi}\int_{R^2}dx^1dx^2\frac{a_2^2}{(r^2+a_2^2)^2}=1.
\end{eqnarray}

Of course, such behaviour is well known from the study of topological solitons \cite{Manton}. The space 
of parameters of the field configuration is not complete (has `holes') and the integrand of the charge
density has corresponding delta functions, which are not seen in (\ref{new}).
 The interesting property here 
is that this process of the vortex shrinking to the delta function and then expanding 
again is a function of time; {\it i.e.} is part of the dynamics of the system
and is described by our solution.

The second and related important fact comes from the study of the integral
 $ \mathcal{I}^{(2)}$. One can check that when the vortex shrinks to the
 delta function ({\it i.e.} the cancellation takes place) the integral 
\begin{eqnarray}
 \mathcal{I}^{(2)}=\frac{1}{\pi}\int_{R^2}dx^1dx^2\frac{a_3^2(k_1-k_2)^2r^2}{(r^2+a_2^2)^2|\vec r-\vec r_n|^2}
\end{eqnarray} diverges. This divergence comes 
from the singularity at the point $\vec r=\vec r_n$
which is responsible for the energy of the solution becoming infinite. 
Clearly,  from a physical point of view such field configurations should be excluded.

\item {Anti-holomorphic solutions}

We can now also apply the transformation (\ref{ortho}) to (\ref{sol})
 and this would give us an
anti-holomorphic solution. Its properties are not very different
 from what we had for the holomorphic
one (except that the choice and meaning of parameters is different) so
 we do not discuss it here.
\end{itemize}

\subsubsection {Further Comments}

In our discussion so far we have assumed that all $y_+$ dependence 
of the 2-dimensional $u(z)$ is of the form of phase factors $\exp(iky_+)$'s.
There is, of course, no need to be so restrictive. We could make the parameters
of the 2-dimensional $u(z)$ depend on $y_+$ in a more general way.
Thus we could consider, for instance, also
\begin{equation}
 u(z,y_+)\,=\, \lambda \frac{1}{z-a(y_+)},
\label{mov}
\end{equation}
where $a(y_+)$ is an arbitrary function.

Then, taking {\it e.g.} $a(y_+) = a\,y_+$ would result in a vortex located at 
$x^2=0$, $x^1=ax^3$ moving in the $x^3$ direction with the velocity of light.
Taking a more complicated function, {\it e.g.} $a(y_+) = a\,y_+^2$ would result
in a curved vortex $x^1=a(x^3)^2$ etc.
One can also combine this dependence, for systems of more vortices,
 with the other dependences discussed above. This complicates the discussion
but does not change its main  features, hence in the remainder of this
paper we return to the discussion of the dependence on $y_+$ through 
the phase factors.

One could naively think that infiniteness of the total energy of our solution is related to some  ``improper'' choice of the dependence on $y_+$. This is not true since the origin of the divergence comes from the topological nature of $\mathcal{H}^{(1)}$. The fact that $\mathcal{H}^{(1)}$ is a total derivative prevents the dependence of  $\mathcal{H}^{(1)}$ on any parameters (including any depending on $y_+$). One can note that for some special cases like $u=z^2\exp{(-a y^2_+)}$ the contribution to the total energy coming from $\mathcal{H}^{(2)}$ is finite but the total energy remains infinite since $\mathcal{H}^{(1)}$ contribution is always present.

\subsection{The $CP^2$ model}
Next we consider solutions of the $CP^2$ model.
 First we look  at the holomorphic ones.

\subsubsection{The holomorphic solutions}

The simplest $CP^2$ model solution can be obtained by adding to 
the system (\ref{sys1}) a constant third function, {\it i.e.} define  
\begin{eqnarray}
f_1(z,y_+)&=&z^2+a_1\,z\,e^{ik_1y_+}\nonumber\\ 
\qquad f_2(z,y_+)&=&a_2\,z+a_3\,e^{ik_2y_+}\nonumber\\ 
\qquad f_3(z,y_+)&=&a_4.\label{sys2}
\end{eqnarray}
Then we can define holomorphic configurations as $u_i=\frac{f_i}{f_3}$, $i=1,2$, i.e.  
\begin{eqnarray}
\label{uicp2}
u_1(z,y_+)=\frac{z^2+a_1\,z\,e^{ik_1y_+}}{a_4},\qquad u_2(z,y_+)=\frac{a_2\,z+a_3\,e^{ik_2y_+}}{a_4}.\label{def1}
\end{eqnarray}
Alternatively, we can interchange $f_2\leftrightarrow f_3$ and consider
the holomorphic configurations 
\begin{eqnarray}
\label{uicp2tilde}
\tilde u_1(z,y_+)=\frac{z^2+a_1\,z\,e^{ik_1y_+}}{a_2\,z+a_3\,e^{ik_2y_+}},\qquad 
\tilde u_2(z,y_+)=\frac{a_4}{a_2\,z+a_3\,e^{ik_2y_+}}.\label{def2}
\end{eqnarray}
Note from (\ref{udef}) that such an interchange corresponds to a phase 
transformation in ${\cal Z}$, so both configurations describe the same solution
of the $CP^2$ model. Note also (easier from (\ref{uicp2tilde})) that
 when $a_4\rightarrow 0$ this $CP^2$ solution reduces to the holomorphic
 $CP^1$ solution discussed before. In fact,
 $\tilde u_2$ vanishes, and $\tilde u_1$ becomes the $CP^1$ $u$-field. 

The integrals $\mathcal{I}^{(1)}$, $\mathcal{I}^{(2)}$ for this $CP^2$ holomorphic
 solution (using definition (\ref{def1}) or (\ref{def2})) now take the form
\begin{eqnarray}
\mathcal{I}^{(1)}=\frac{1}{\pi}\int_{R^2}dx^1dx^2\frac{\mathcal{A}}{\mathcal{C}^2},\qquad \qquad 
\mathcal{I}^{(2)}=\frac{1}{\pi}\int_{R^2}dx^1dx^2\frac{\mathcal{B}}{\mathcal{C}^2}\label{integrals3}
\end{eqnarray}
where $\mathcal{A}$, $\mathcal{B}$, $\mathcal{C}$ differ from $A$, $B$, $C$ given
 by (\ref{A}), (\ref{B}) and (\ref{C}) by terms proportional to $a_4^2$, {\it i.e.} 
\begin{eqnarray}
\mathcal{A}&=&A+a_4^2[a_1^2+a_2^2+4r^2+4a_1r\cos{(\varphi-k_1y_+)}]\\
\mathcal{B}&=&B+a_4^2[a_1^2k_1^2r^2+a_3^2k_2^2]\\
\mathcal{C}&=&C+a_4^2.
\end{eqnarray}
The Hamiltonian density $\mathcal{H}^{(2)}$, which is proportional 
to $\frac{\mathcal{B}}{\mathcal{C}^2}$, is  now regular at $\vec{r}=\vec r_n$ 
and $y_+=\xi_n$ for $a_1=\frac{a_3}{a_2}$ (where previously we had a singularity)
 as now it takes the value
$$
\left.\frac{\mathcal{B}}{\mathcal{C}^2}\right|_{\vec r=\vec r_n,\, y_+=\xi_n}=\frac{a_3^2}{a_4^2}\left[\frac{a_3^2}{a_2^4}k_1^2+k_2^2\right].
$$
Hence we note that going to the $CP^{2}$ manifold (by taking $a_4\ne0$) 
has `filled in the hole' in the space
of parameters ({\it i.e.} as the system evolves none of its vortices shrinks to 
the delta function).

Note also that the energy density is independent of $y_+$ 
in four cases: $k_1=k_2$, $a_1=0$, $a_2=0$ and $a_3=0$.

\subsubsection{The mixed solution}

Next we look at the `new' mixed solutions. First we use (\ref{ortho}) 
to calculate $P_z f$. 
We find that for the system  (\ref{sys2}) they take the form

\begin{eqnarray}
P_zf_1&=&a_4^2e^{ik_1y_+}\left[2ze^{-ik_1y_+}+a_1\right]\nonumber\\
&+&e^{ik_1y_+}\left[a_3+a_2\bar{z}e^{ik_2y_+}\right]\left[a_1a_3+2a_3ze^{-ik_1y_+}+a_2z^2e^{-i(k_1+k_2)y_+}\right]\nonumber\\
P_zf_2&=&a_2a_4^2\nonumber\\
&-&e^{ik_2y_+}\bar{z}\left[a_1+\bar{z}e^{ik_1y_+}\right]\left[a_1a_3+2a_3ze^{-ik_1y_+}+a_2z^2e^{-i(k_1+k_2)y_+}\right]\nonumber\\
P_zf_3&=&-a_4e^{-ik_2y_+}\left[a_2a_3+a_2^2\bar{z}e^{ik_2y_+}+\bar{z}e^{ik_2y_+}(\bar{z}e^{ik_1y_+}+a_1)(2ze^{-ik_1y_+}+a_1)\right]\nonumber
\end{eqnarray}

When written in terms of $u_i$ this mixed solution is given by
\begin{eqnarray}
u_1(z,\bar{z},y_+)=\frac{P_zf_1}{P_zf_3},\qquad u_2(z,\bar{z},y_+)=\frac{P_zf_2}{P_zf_3}.\label{solmix}
\end{eqnarray}

Note that in the limit $a_4\rightarrow 0$ the mixed solution
 (\ref{solmix}) becomes the anti-holomorphic solution of the $CP^1$ model
mentioned before. However for $a_4\ne0$ the solution is different.
This time the expressions for the energy density 
are quite complicated - so we do not present them here.
However, we note that to guarantee the convergence of the integral $\mathcal{I}^{(2)}$ 
we have to require that $a_2\neq 0$.

To demonstrate that the energy per unit length does
 not depend on $y_+$ can be checked without much effort.
First, we observe that the overall factors $e^{ik_jy_+}$ in $P_zf_k$ do 
not matter as they cancel in the expressions for $|u_j|^2$ and 
for $|\Delta\cdot u_j|^2$. Hence, the only relevant  expressions 
are of the form $ze^{-ik_jy_+}=re^{i(\varphi-k_jy_+)}$ 
and $\bar{z}e^{ik_jy_+}=re^{-i(\varphi-k_jy_+)}$.
 When $k_1=k_2\equiv k$ the energy density depends only on the 
combination $(\varphi-ky_+)$ and $r$ showing that the only effect of
the dependence on time is a rotation and, in consequence, the independence
 of the energy per unit length on $y_+$ (or $x^0$ for given $x^3$).
 The other cases guaranteeing this  are  $a_1=0$ and  $a_3=0$.

\subsubsection{The anti-holomorphic solution}

Finally we look at the corresponding anti-holomorphic solution. 
Such a solution derived from the system (\ref{sys2}) takes the form
\begin{eqnarray}
u_1(\bar z,y_+)&=&\frac{P_z^2f_1}{P_z^2f_3}=\frac{a_2a_4e^{i(k_1+k_2)y_+}}{a_1a_3+\bar z e^{ik_1y_+}(2a_3+a_2\bar z e^{ik_2y_+})}
\\
u_2(\bar z,y_+)&=&\frac{P_z^2f_2}{P_z^2f_3}=-\frac{a_4e^{ik_2y_+}(a_1+2\bar z e^{ik_1y_+})}{a_1a_3+\bar z e^{ik_1y_+}(2a_3+a_2\bar z e^{ik_2y_+})}.
\end{eqnarray}
Note that, like for the `mixed case', we have to require that $a_2\ne 0$ as otherwise 
$$
\mathcal{H}^{(1)}=0,\qquad \mathcal{H}^{(2)}=8\pi M^2\frac{k_2^2a_3^2a_4^2}{(a_3^2+a_4^2)^2}.
$$

In the next subsection we will produce an explicit example of these field 
configurations and discuss some of their
properties. To avoid the problems mentioned above our example will have $a_2\neq 0$. 
Note that in such a case
 the conditions of the independence of the energy per unit length on $y_+$
 are the same as for the mixed solution.

\subsubsection{An example}

\begin{figure}
\begin{center}
\includegraphics[width=0.3\textwidth]{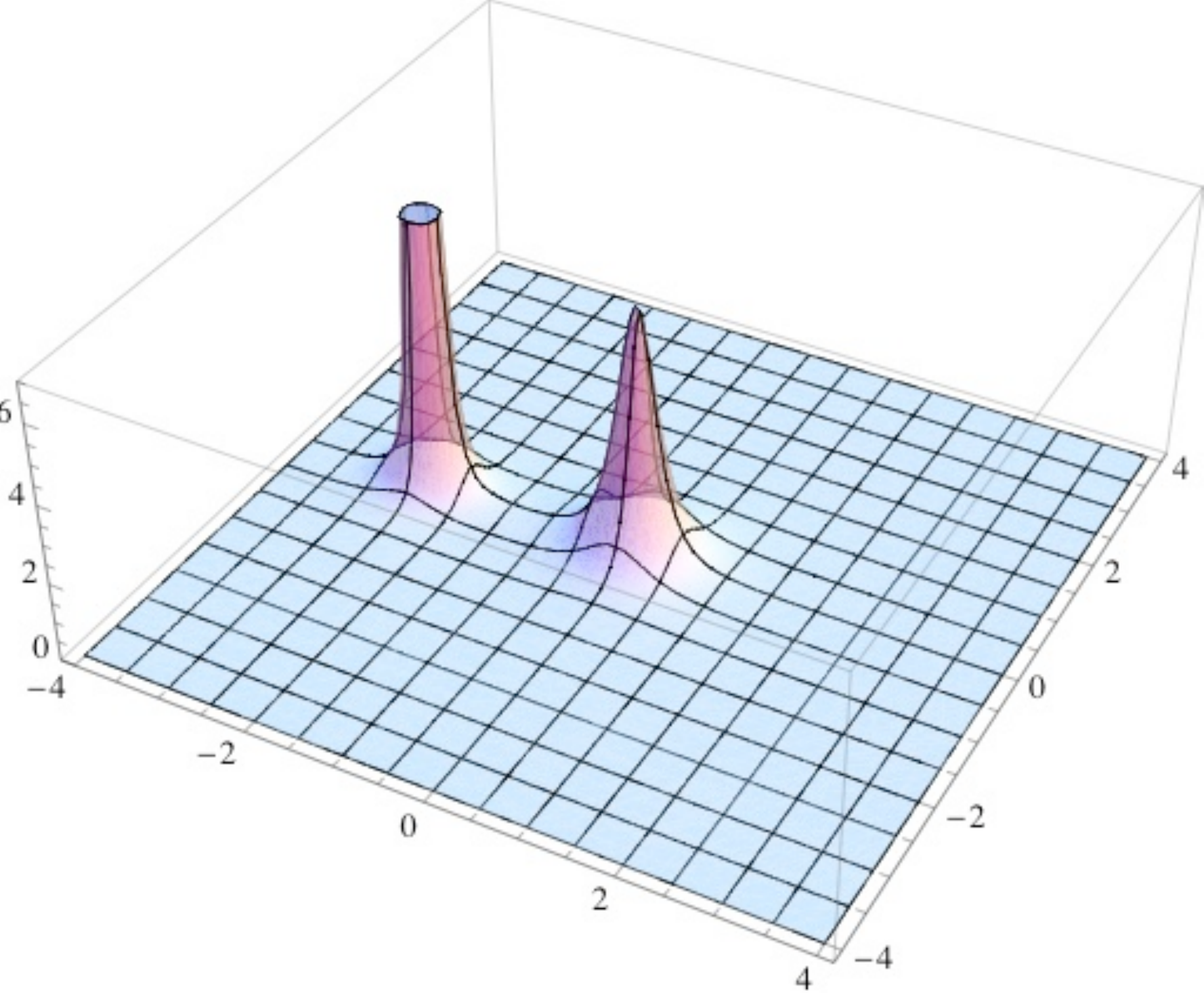}\includegraphics[width=0.3\textwidth]{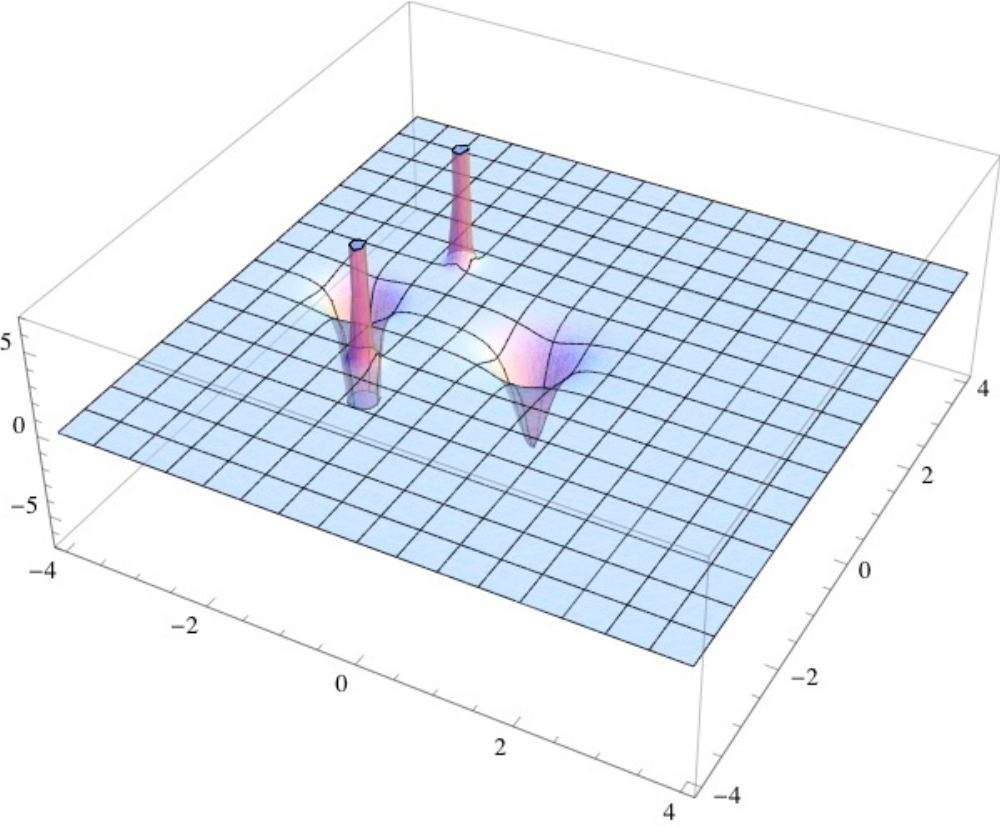}\includegraphics[width=0.3\textwidth]{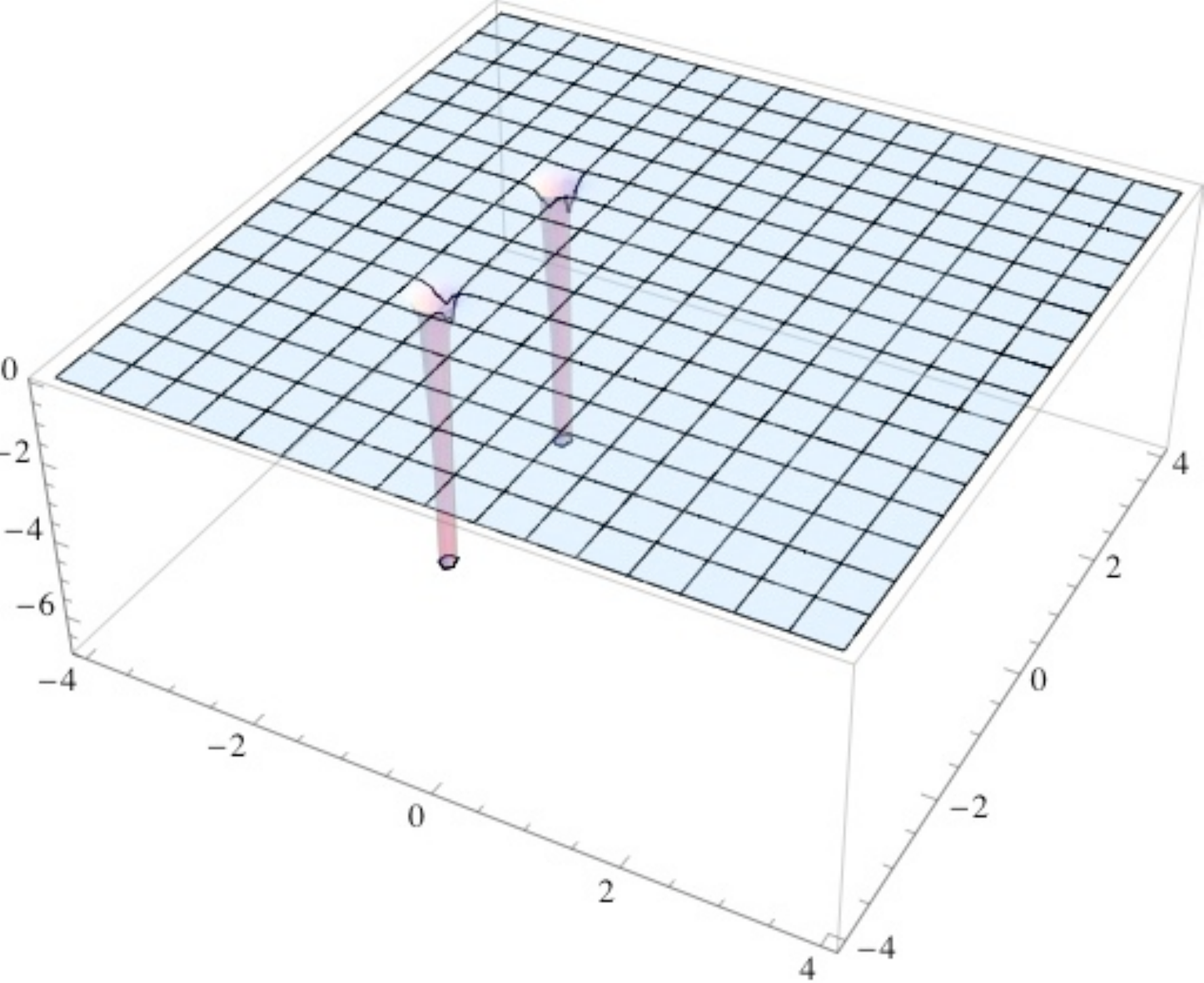}
\caption{The functions $\pi\rho_{\rm top}$ (where $\rho_{\rm top}$ is a topological charge density) for $x^0=0$, $x^3=0$, $a_1=2.5$, $a_2=0.6$, $a_3=1.0$, $a_4=0.01$, $k_1=1$ and $k_2=2$. The left picture corresponds to the holomorphic solution, the central picture corresponds to the mixed solution and the right picture to the anti-holomorphic solution.}
\label{top}
\end{center}
\end{figure}

\begin{figure}
\begin{center}
\includegraphics[width=0.3\textwidth]{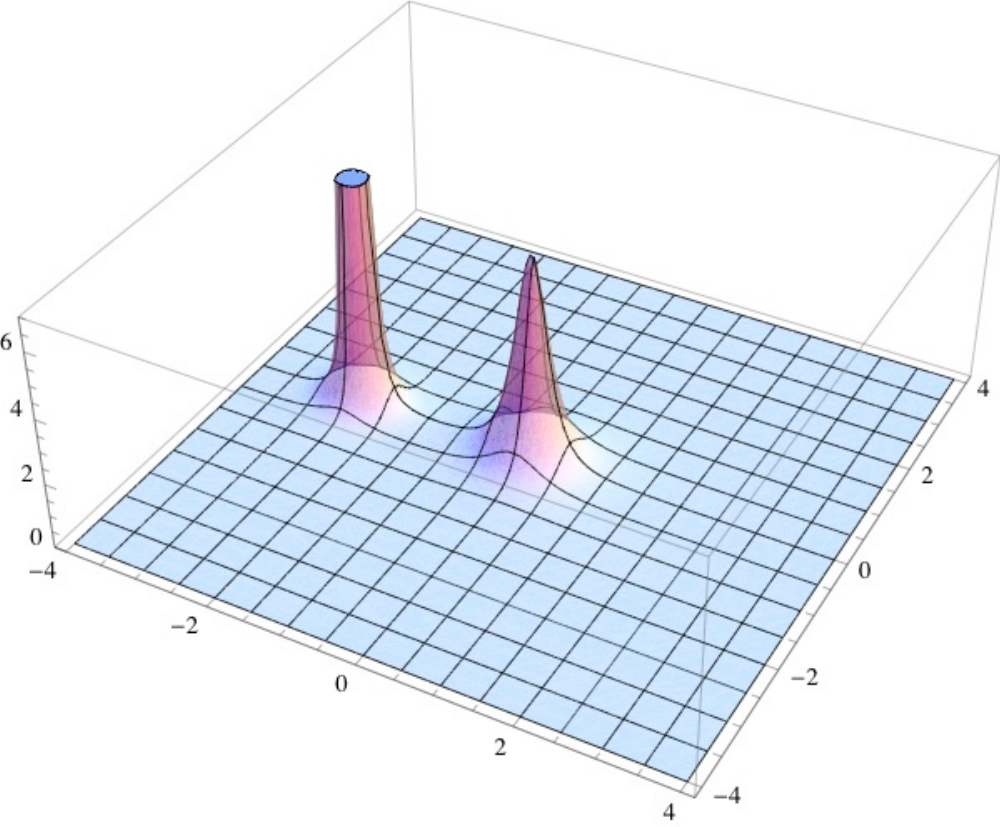}\includegraphics[width=0.3\textwidth]{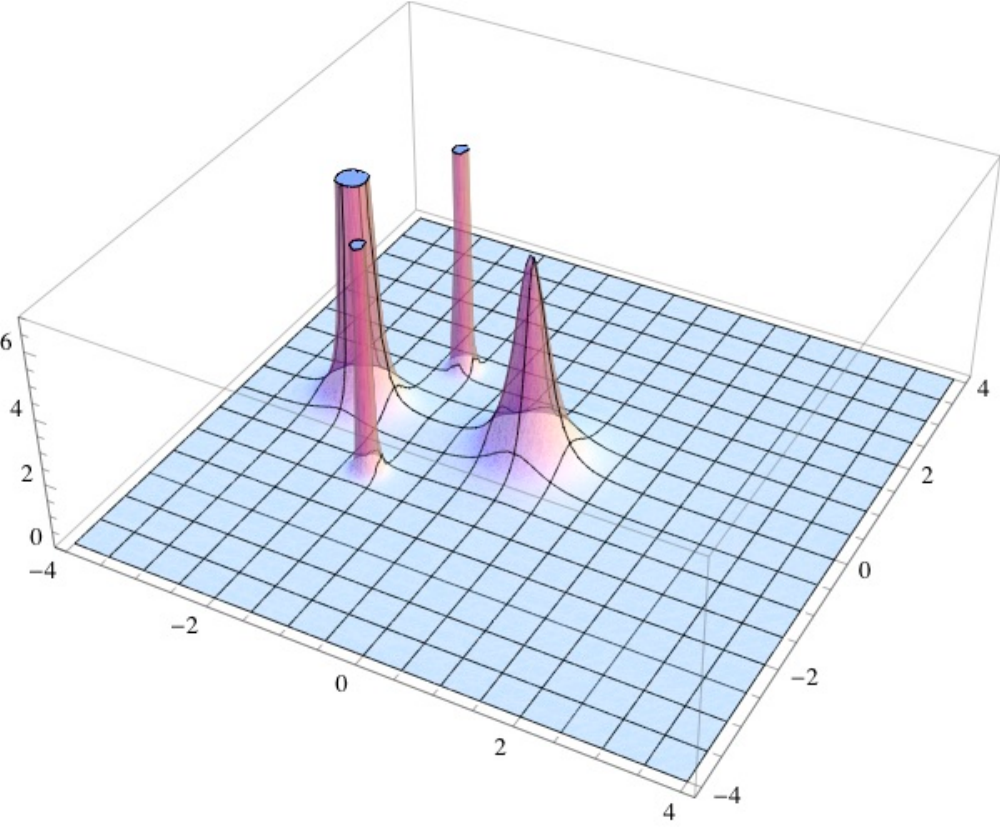}\includegraphics[width=0.3\textwidth]{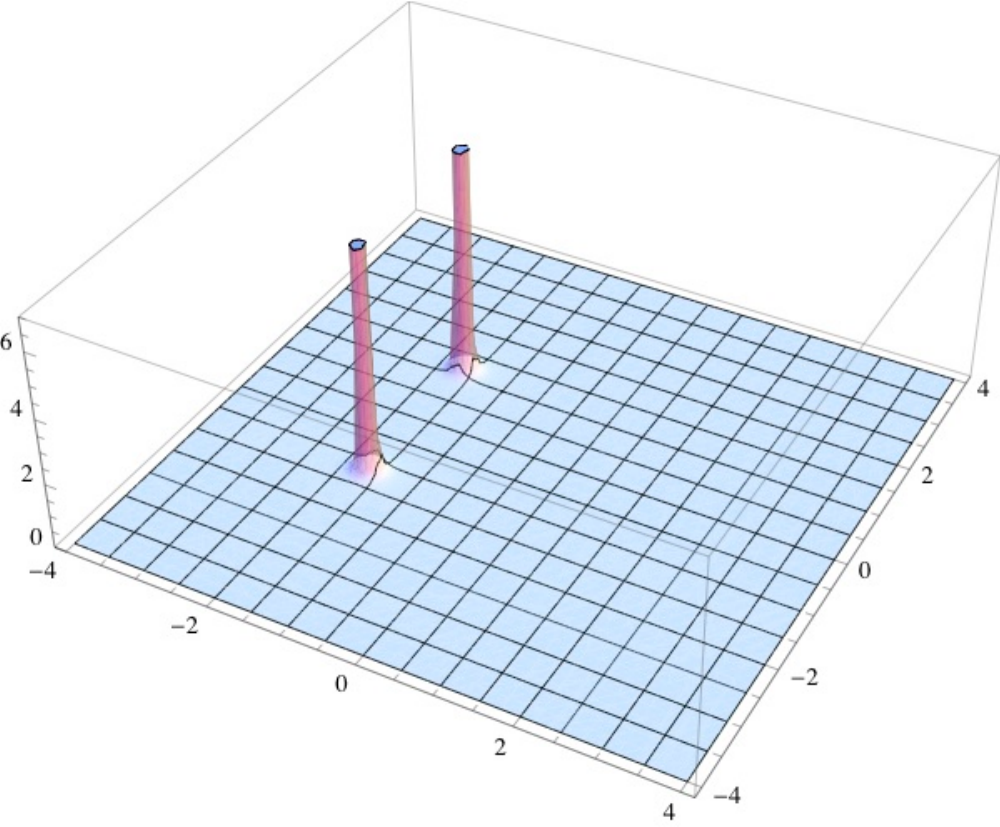}
\caption{The functions $\mathcal{H}^{(1)}/8M^2$ (proportional to topological contribution to the energy density) for $x^0=0$, $x^3=0$, $a_1=2.5$, $a_2=0.6$, $a_3=1.0$, $a_4=0.01$, $k_1=1$ and $k_2=2$. The left picture corresponds to the holomorphic solution, the central picture corresponds to the mixed solution and the right picture to the anti-holomorphic solution.}
\label{en1}
\end{center}
\end{figure}

\begin{figure}
\begin{center}
\includegraphics[width=0.3\textwidth]{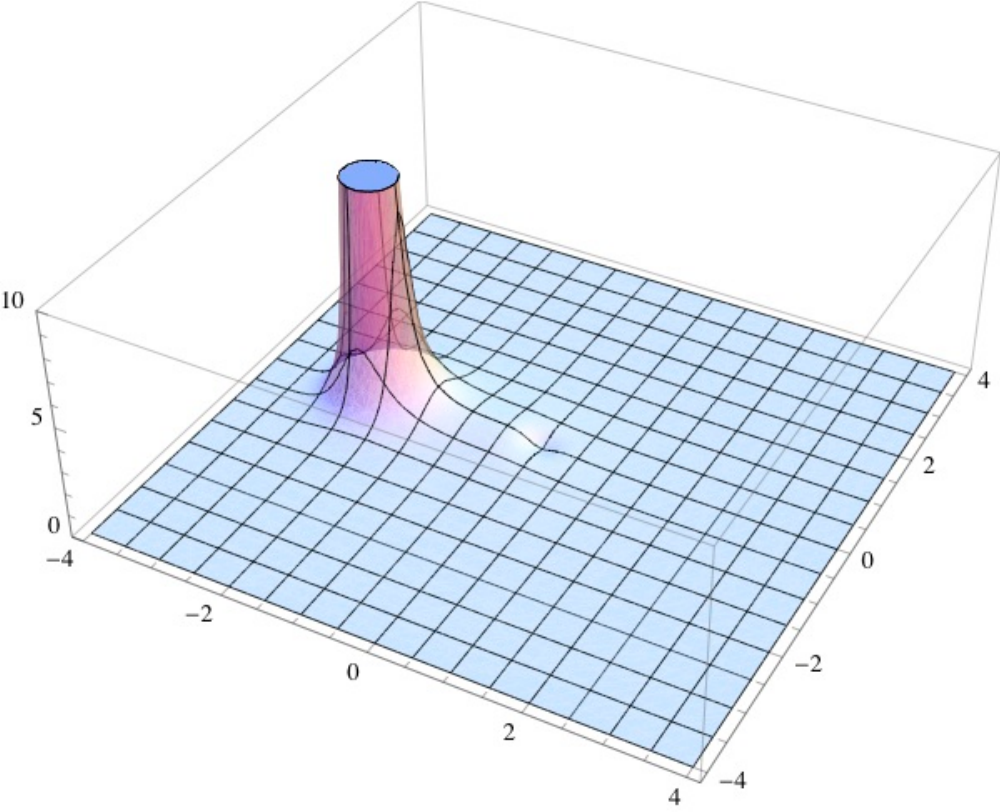}\includegraphics[width=0.3\textwidth]{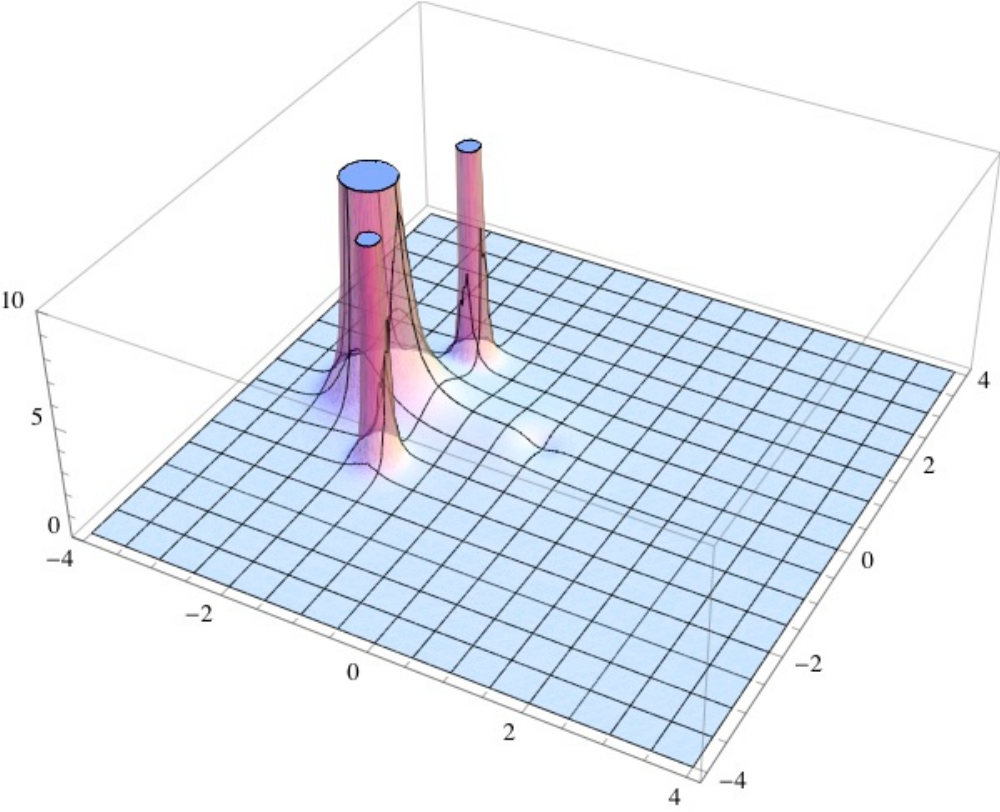}\includegraphics[width=0.3\textwidth]{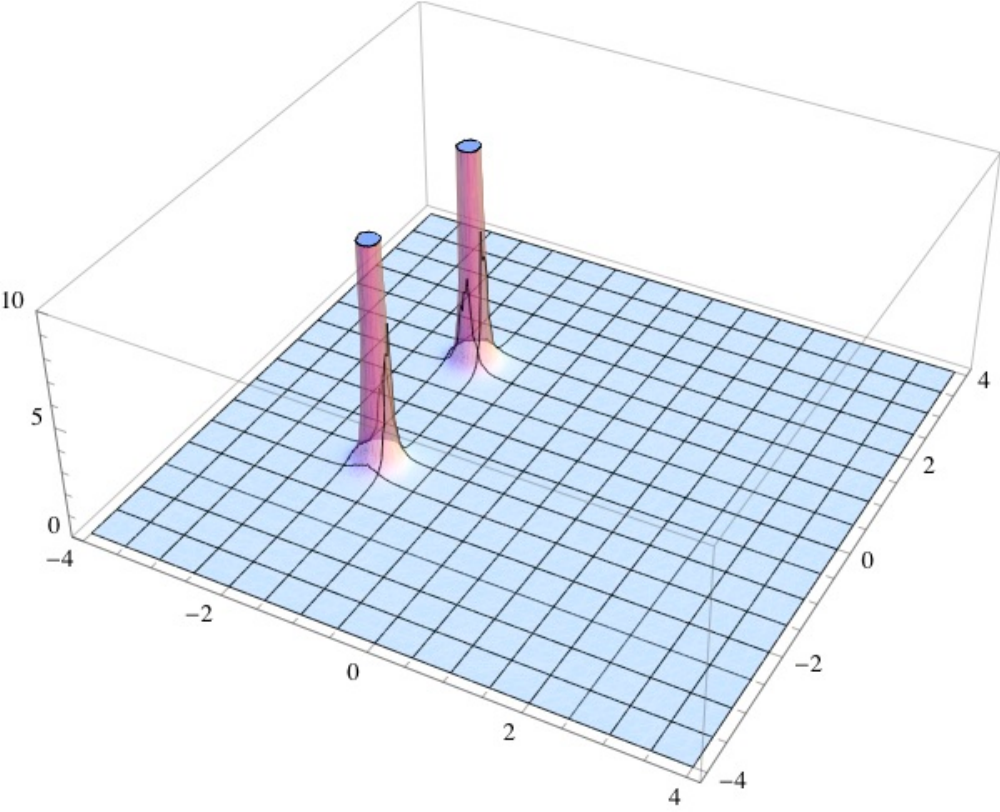}
\caption{The functions $\mathcal{H}^{(2)}/8M^2$ (proportional to wave contribution to the energy density) for $x^0=0$, $x^3=0$, $a_1=2.5$, $a_2=0.6$, $a_3=1.0$, $a_4=0.01$, $k_1=1$ and $k_2=2$. The left picture corresponds to the holomorphic solution, the central picture corresponds to the mixed solution and the right picture to the anti-holomorphic solution.}
\label{en2}
\end{center}
\end{figure}

In our example we start with 
the set of functions (\ref{sys2}) for which we have chosen the following 
values of parameters: $a_1=2.5$, $a_2=0.6$, $a_3=1.0$, $a_4=0.01$ $k_1=1.0$ 
and $k_2=2.0$. The topological charge of the holomorphic solution is 
then $Q_{\rm top}=2$. The topological charge density at $x^0=0$
 (and for $x^3=0$) has two peaks - one of them is localized at $z=0$, 
the other a bit further out - see Fig. \ref{top}.
 For the holomorphic solution the topological charge density 
is proportional to the energy density and this leads to the energy per unit length 
being given by $8\pi M^2 \mathcal{I}^{(1)}$.
 The integrand $\mathcal{H}^{(1)}/8M^2$ is sketched in Fig. \ref{en1}.
 The contribution coming from the waves $\mathcal{H}^{(2)}/8M^2$
 is plotted in Fig. \ref{en2}. The mixed solution
 generated by the application of the $P_z$ operator
according to (\ref{ortho}) leads to a solution which has $Q_{\rm top}=2-2=0$ 
and $\mathcal{I}^{(1)}=2+2=4$. As is easy to see from Fig. \ref{top}
 the application of $P_z$ has changed two holomorphic peaks
 into two anti-peaks and in addition it has generated two new peaks.
 The energy density  $\mathcal{H}^{(1)}$ thus has four peaks and
 $\mathcal{H}^{(2)}$ only three (with the zero in the place of the 
fourth $\mathcal{H}^{(1)}$ one).

\begin{figure}
\begin{center}
\includegraphics[width=0.3\textwidth]{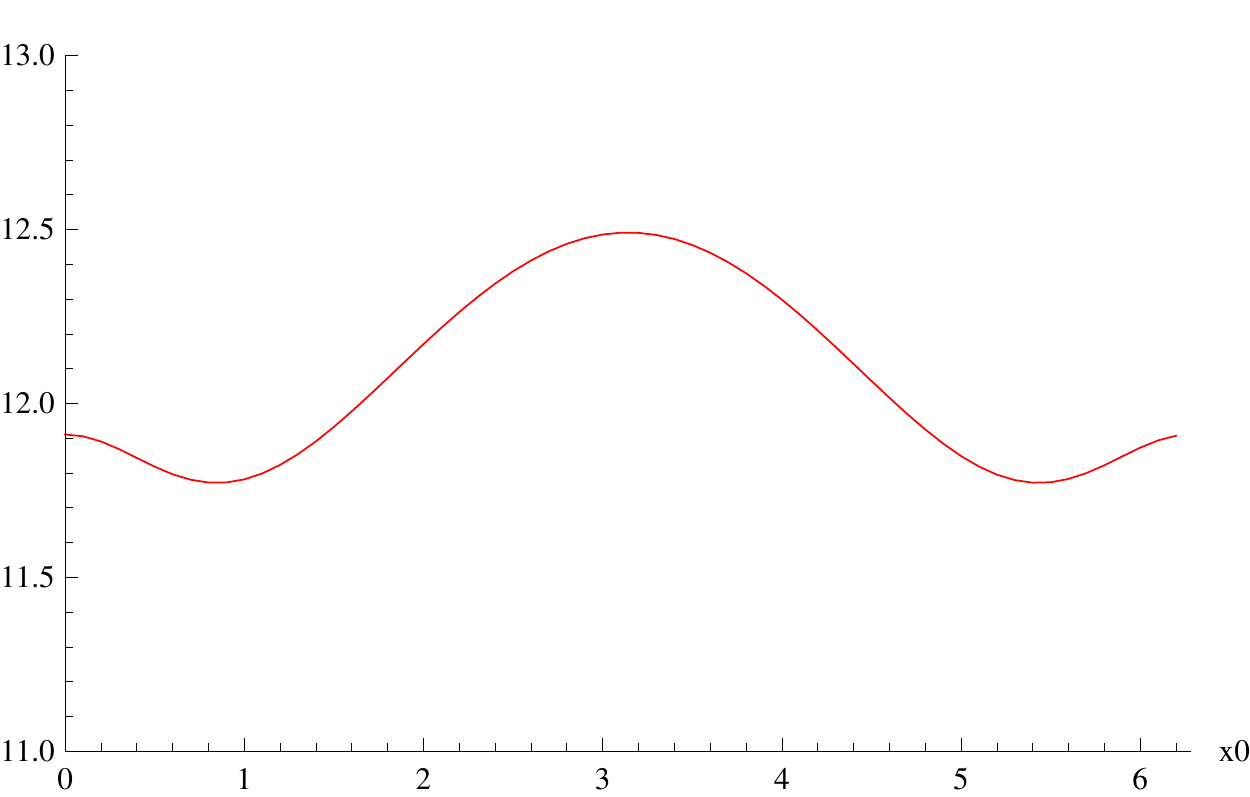}\includegraphics[width=0.3\textwidth]{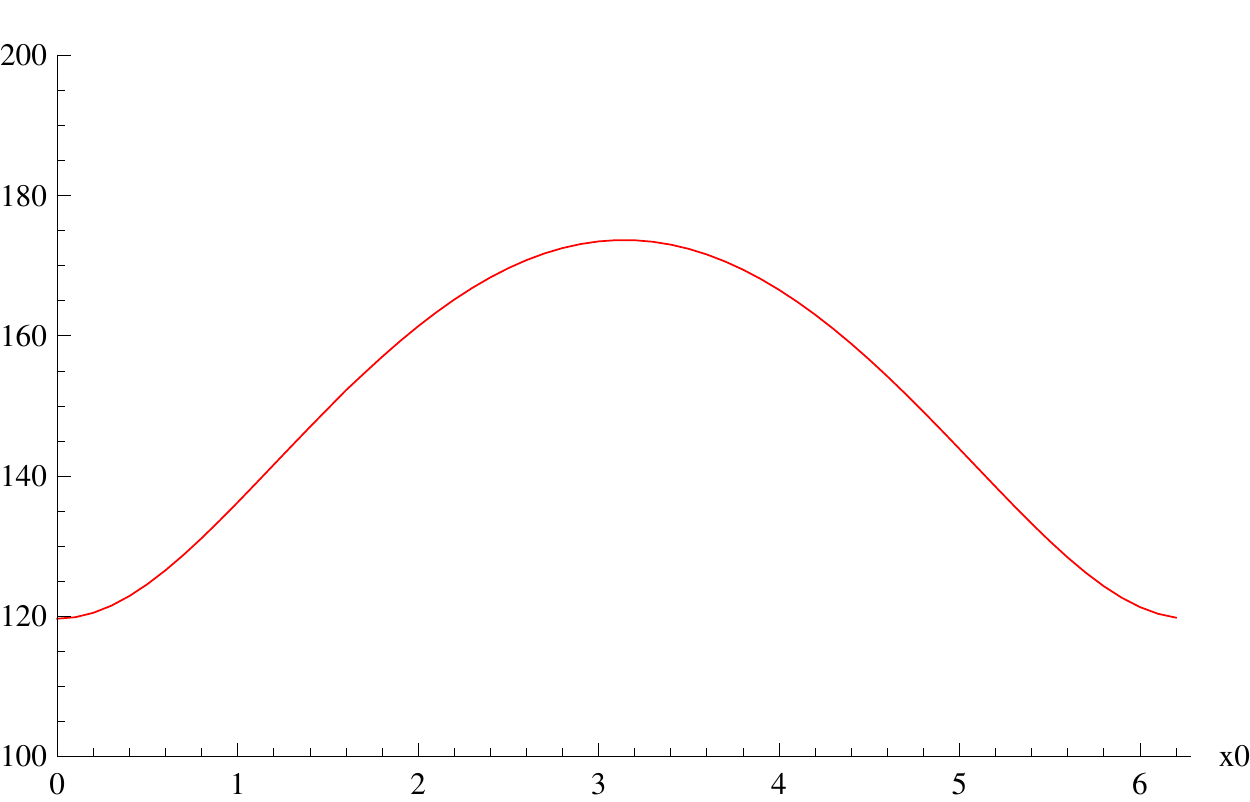}\includegraphics[width=0.3\textwidth]{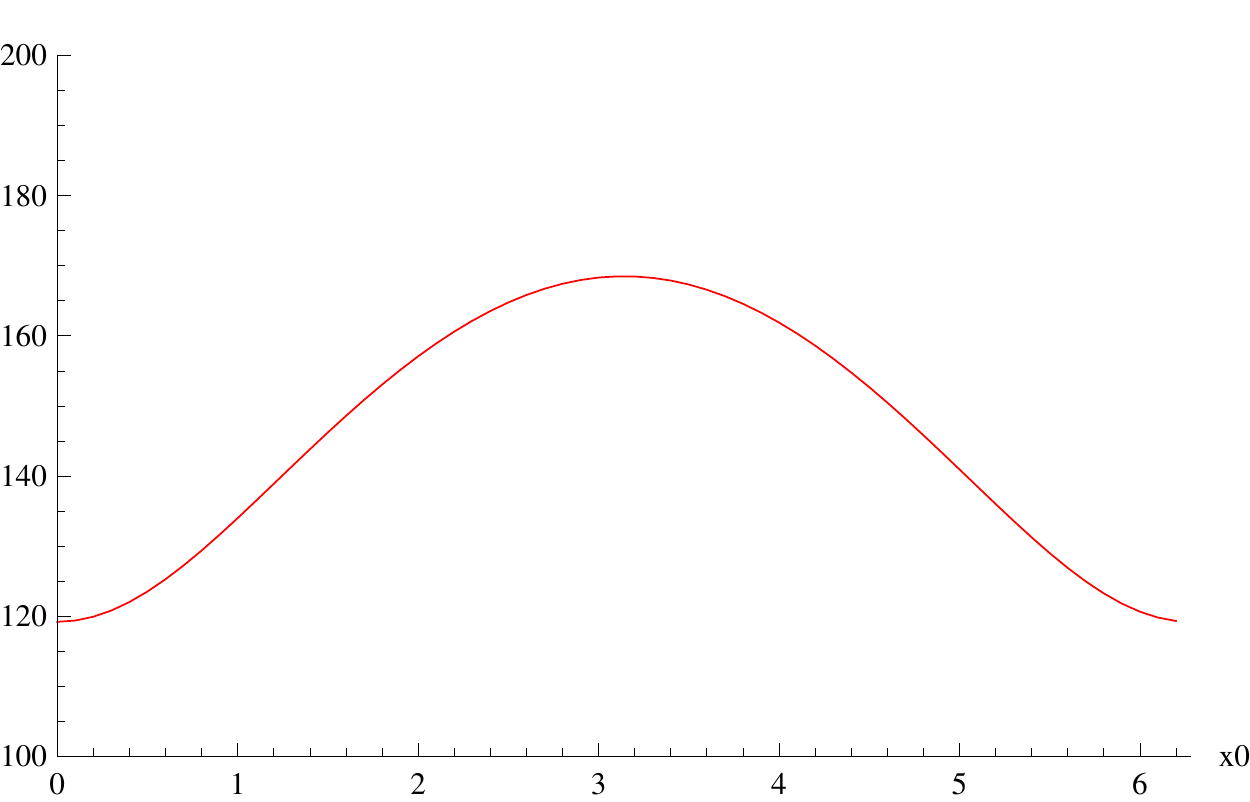}
\caption{The integral $\mathcal{I}^{(2)}$ as the function of $x^0\in[0,2\pi]$. The other parameters read: $x^3=0$, $a_1=2.5$, $a_2=0.6$, $a_3=1.0$, $a_4=0.01$, $k_1=1$ and $k_2=2$. The left picture corresponds to the holomorphic solution, the central picture corresponds to the mixed solution and the right picture to the anti-holomorphic solution.}
\label{energy}
\end{center}
\end{figure}

\begin{figure}
\begin{center}
\includegraphics[width=0.3\textwidth]{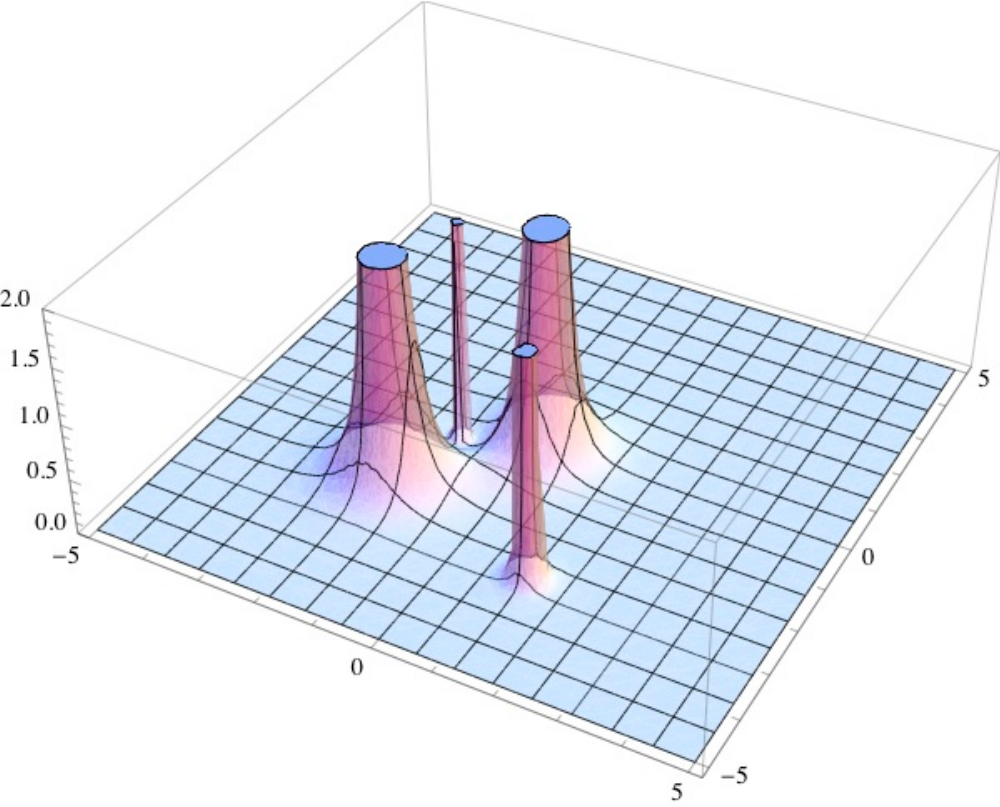}\includegraphics[width=0.3\textwidth]{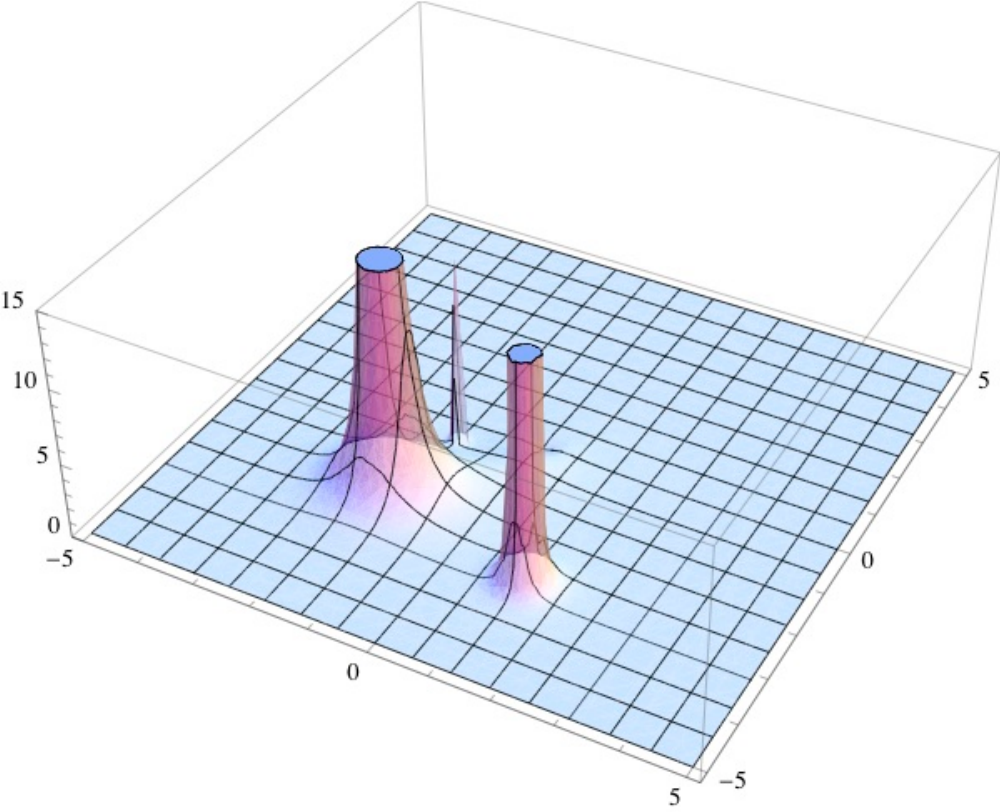}
\includegraphics[width=0.3\textwidth]{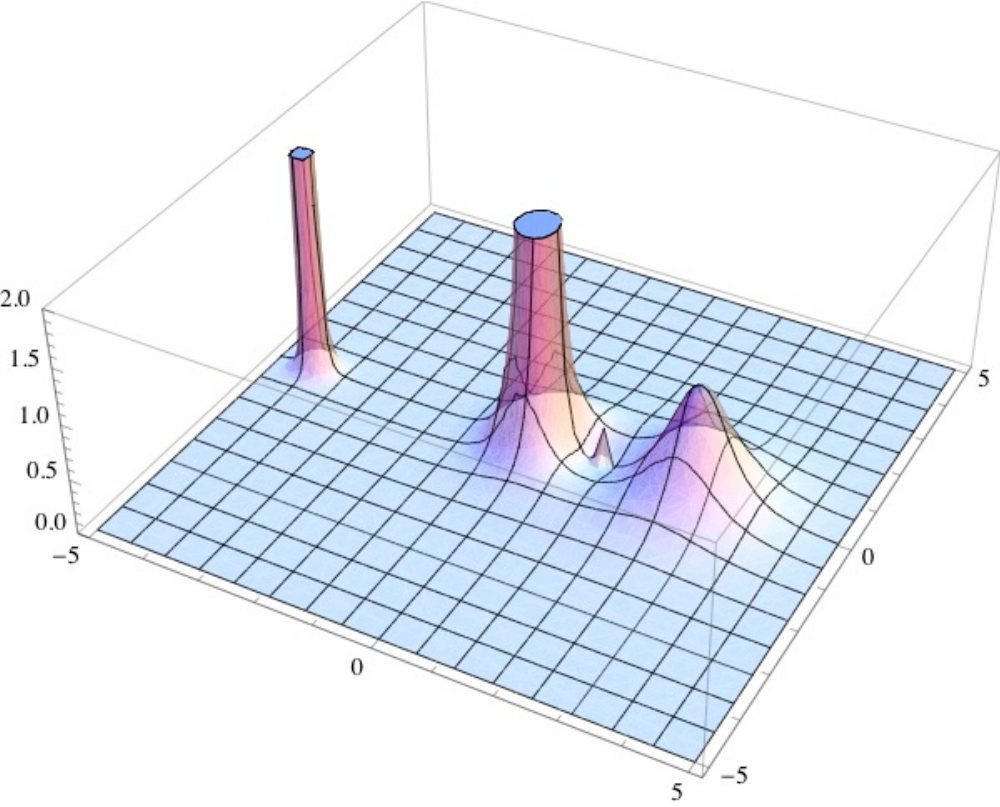}\includegraphics[width=0.3\textwidth]{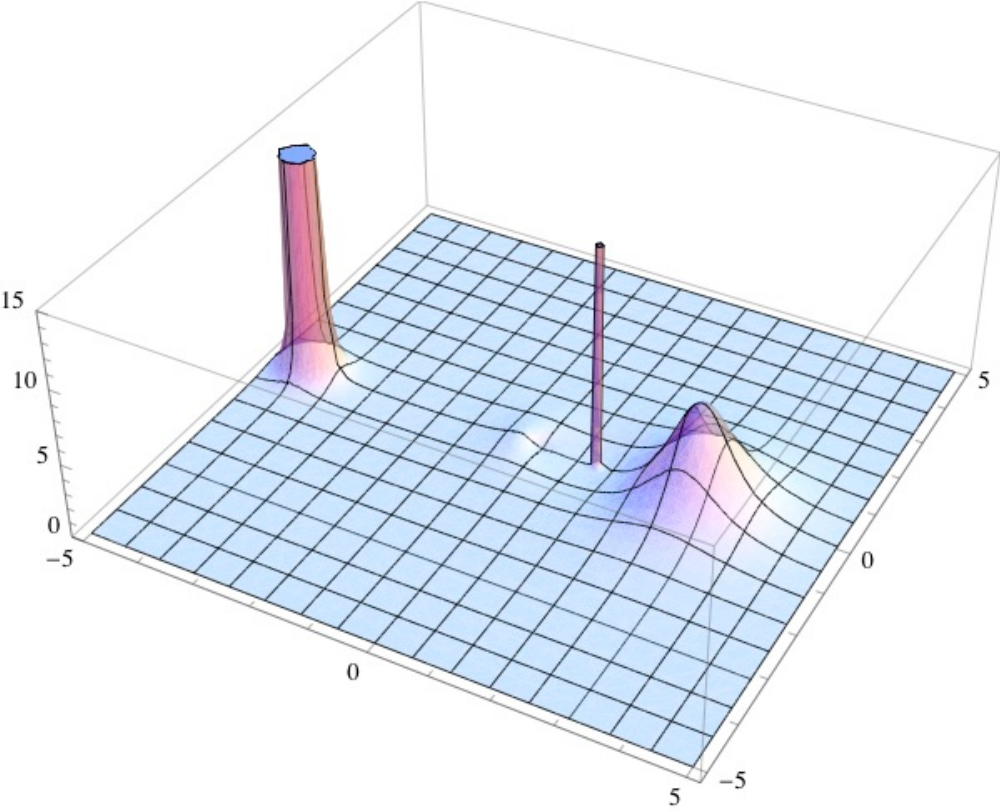}
\includegraphics[width=0.3\textwidth]{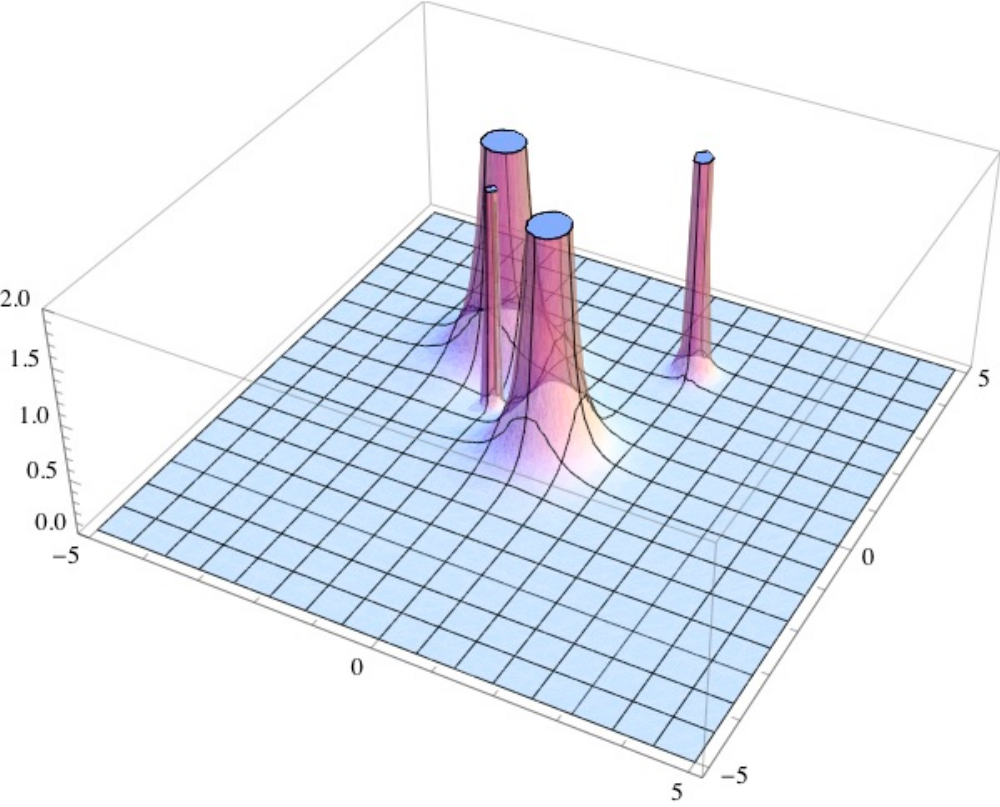}\includegraphics[width=0.3\textwidth]{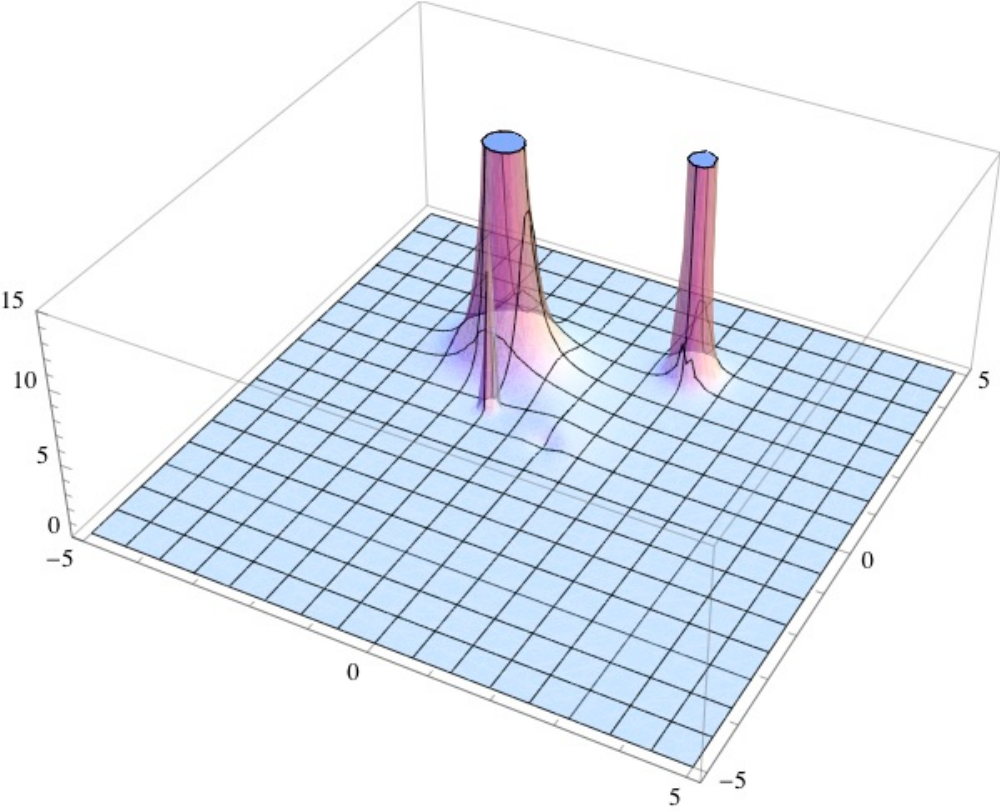}
\caption{Time evolution of the mixed solution for $a_1=2.5$, $a_2=0.6$, $a_3=1.0$, $a_4=0.01$, $k_1=1$ and $k_2=2$. The functions $\mathcal{H}^{(1)}/8M^2$ (left column) and  $\mathcal{H}^{(2)}/8M^2$ (right column) have been considered for $x^3=0$ at the moments $x^0=\pi/4$ (first row), $x^0=\pi$ (second row), $x^0=7\pi/4$ (third row). Their values for $x^0=0$ are sketched at the central pictures of Fig \ref{en1} and Fig \ref{en2}.}
\label{evolution}
\end{center}
\end{figure}

 The next application of the $P_z$ operator  changes two peaks 
 of the topological charge density into two anti-peaks and annihilates 
the previous anti-peaks. Thus the  anti-holomorphic solution is 
characterized by $Q_{\rm top}=-2$ and $\mathcal{I}^{(1)}=2$.
 The contribution to the energy per unit length $8M^2\mathcal{I}^{(1)}$ 
is the same as for the initial (holomorphic) case. 
 Nevertheless, the total energies per unit length for these two
 solutions differ since for solutions of the $CP^2$ model the integrals 
$\mathcal{I}^{(2)}$ are different ($\mathcal{I}^{(2)}_{\rm hol}\neq
 \mathcal{I}^{(2)}_{\rm anti-hol}$).
Let us note that our case has a time-dependent energy
 per unit length (calculated by the integration over the $x^1x^2$ plane). It implies that the dependence of the energy density on  $y_+$ is highly nontrivial. However, the energy per unit length is a periodic function of $y_+$. Only for some special cases, like $k_1=k_2$ etc. the energy per unit length is constant and so does not depend on $y_+$. The time dependence of the energy density for all three solutions is shown in Fig. \ref{energy}. The energy density for the mixed solution for $x^3=0$ and $x_0=\pi/4$, $x_0=\pi$, $x_0=7\pi/4$ is plotted in Fig. \ref{evolution}. For the case $x^0=\pi$ the peaks are maximally separated (this is not very clear without a detailed study of some other values $x^0$). In this case the energy takes its maximal value, see Fig. \ref{energy}.

\begin{figure}
\begin{center}
\includegraphics[width=0.5\textwidth]{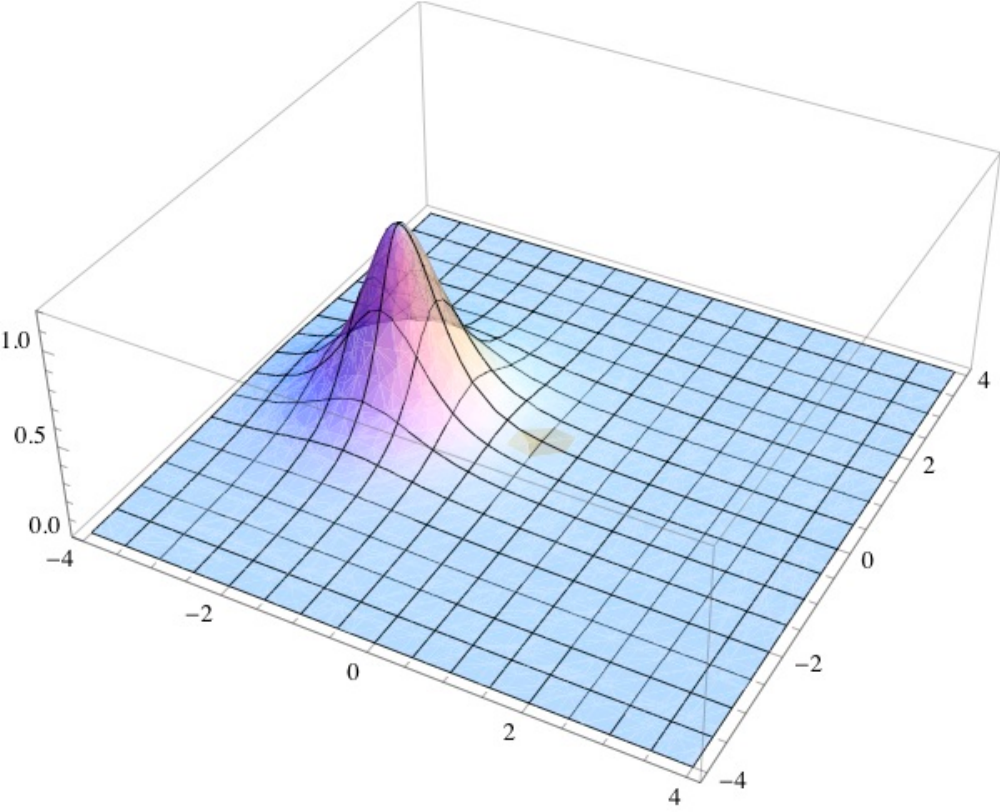}\includegraphics[width=0.5\textwidth]{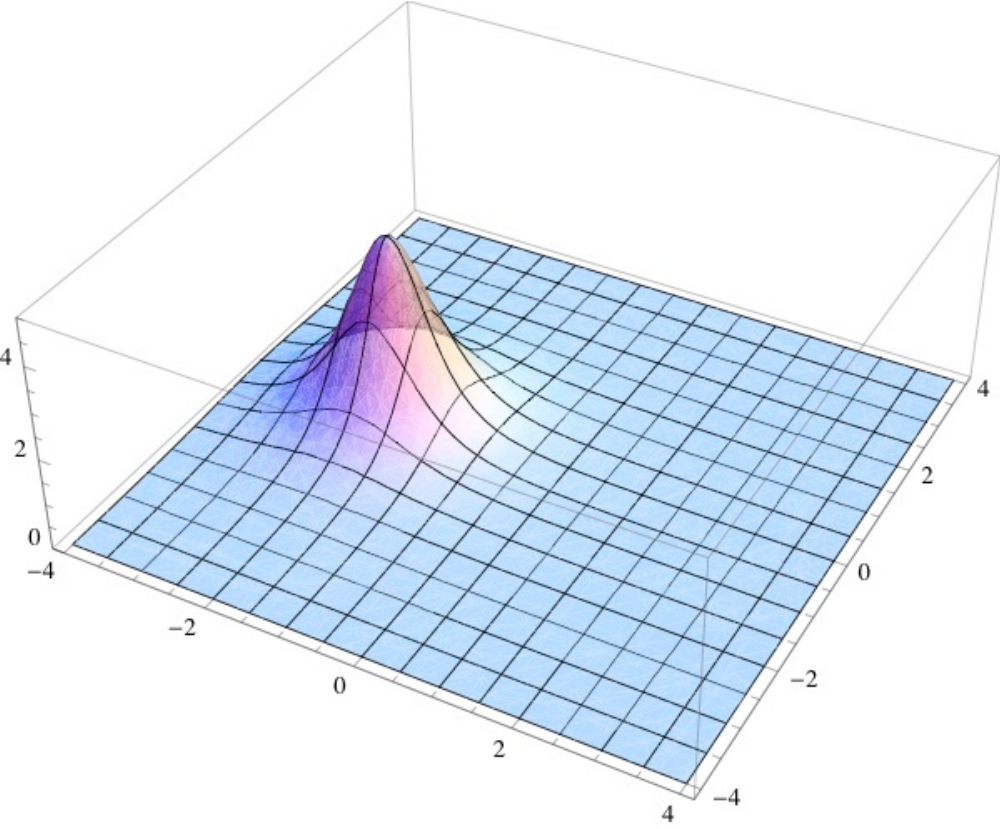}
\caption{The energy density of the spiral solution - the topological part (left) and the wave part (right).  Here $a_1=2$, $a_2=1$, $a_3=0$, and $x^0=0$, $x^3=0$ and $k_1=1$. The maxima of the energy density for both contributions are located at the same point on the plane $x^1x^2$ corresponding with the minimum of $n^3$ (see Fig \ref{spiral_isovector}); here $x^1=-2$, $x^2=0$.}
\label{spiral_ener}
\end{center}
\end{figure}

\section{Conclusions and Further Comments}

In this paper we have demonstrated that the $CP^N$ model in (3+1)
dimensions has many classical solutions. Our construction has been based on the 
observation that one can generalise ideas used in the construction
of solutions of the $CP^N$ model in (2+0) dimensions
and generate vortex and vortex-antivortex like solutions of this model
in (3+1) dimensions.
Like for the model in (2+0) dimensions we can generate these solutions 
from field configurations described by polynomial functions of $x^1+i\epsilon_1 x^2$.
This time the coefficients of these functions could be also functions of
$x^3+\varepsilon_2\, x_0$. The energy of 
such configurations is infinite (as the energy density is independent
of $x^3$) and so we interpret these solutions as describing systems of vortices  and
antivortices.

Of course our expressions solve equations in (3+1) dimensions
and they also determine the dynamics of these vortices.

In this paper we have only looked at the simplest solutions (corresponding
to very few vortices) with the time dependence being described by simple 
phase factors. Even in this case the observed dynamics is quite complicated
and has exhibited various interesting properties.
In particular, we have shown that the vortices can rotate in space 
(physical and internal) and their energy per unit length
of the vortex can vary in time. During this time evolution
some vortices can shrink to delta functions and then expand again
often being characterised by a very periodical behaviour.

 One other unusual property is their dependence on the
distance between the vortices: the energy density of
two vortices can depend on the distance
between them and can possess a minimum at a specific 
value of this distance. This suggests that vortices which are located
at non-minimal distances may be unstable and so could try to reduce their 
energy per unit length by moving
towards this optimal  
configurations. However, their configurations are solutions for any
distance as their infinite  
`inertia' stops them from moving towards each other without an external push.

We are now looking at other properties of these and other solutions.

{\bf Acknowlegment:} L.A. Ferreira and W.J. Zakrzewski would like to thank
the Royal Society (UK)
for a grant that helped them in carrying out this work. L.A. Ferreira
is partially supported by CNPq (Brazil) and P. Klimas is supported by
FAPESP (Brazil).

\end{document}